\documentclass[
reprint,
superscriptaddress,
amsmath,amssymb,
aps,
prb,
floatfix
]{revtex4-2}

\usepackage{graphicx}% Include figure files
\usepackage{dcolumn}% Align table columns on decimal point
\usepackage{bm}% bold math
\usepackage[colorlinks,citecolor=magenta]{hyperref}

\begin{document}

\title{Lasing of Moiré Trapped MoSe$_2$/WSe$_2$ Interlayer Excitons Coupled to a Nanocavity}

\author{Chenjiang Qian}
\email{chenjiang.qian@iphy.ac.cn}
\affiliation{Walter Schottky Institut and TUM School of Natural Science, Technische Universit{\" a}t M{\" u}nchen, Am Coulombwall 4, 85748 Garching, Germany}
\affiliation{Beijing National Laboratory for Condensed Matter Physics, Institute of Physics, Chinese Academy of Sciences, Beijing 100190, China}
\affiliation{School of Physical Sciences, University of Chinese Academy of Sciences, Beijing 100049, China}
\author{Mirco Troue}
\affiliation{Walter Schottky Institut and TUM School of Natural Science, Technische Universit{\" a}t M{\" u}nchen, Am Coulombwall 4, 85748 Garching, Germany}
\affiliation{Munich Center for Quantum Science and Technology (MCQST), Schellingstr. 4, 80799 Munich, Germany}
\author{Johannes Figueiredo}
\affiliation{Walter Schottky Institut and TUM School of Natural Science, Technische Universit{\" a}t M{\" u}nchen, Am Coulombwall 4, 85748 Garching, Germany}
\affiliation{Munich Center for Quantum Science and Technology (MCQST), Schellingstr. 4, 80799 Munich, Germany}
\author{Pedro Soubelet}
\affiliation{Walter Schottky Institut and TUM School of Natural Science, Technische Universit{\" a}t M{\" u}nchen, Am Coulombwall 4, 85748 Garching, Germany}
\author{Viviana Villafañe}
\affiliation{Walter Schottky Institut and TUM School of Natural Science, Technische Universit{\" a}t M{\" u}nchen, Am Coulombwall 4, 85748 Garching, Germany}
\author{Johannes Beierlein}
\affiliation{Julius-Maximilians-Universität Würzburg, Physikalisches Institut and Würzburg-Dresden Cluster of Excellence ct.qmat, Lehrstuhl für Technische Physik, Am Hubland, 97074 Würzburg, Germany}
\author{Sebastian Klembt}
\affiliation{Julius-Maximilians-Universität Würzburg, Physikalisches Institut and Würzburg-Dresden Cluster of Excellence ct.qmat, Lehrstuhl für Technische Physik, Am Hubland, 97074 Würzburg, Germany}
\author{Andreas V. Stier}
\affiliation{Walter Schottky Institut and TUM School of Natural Science, Technische Universit{\" a}t M{\" u}nchen, Am Coulombwall 4, 85748 Garching, Germany}
\author{Sven Höfling}
\affiliation{Julius-Maximilians-Universität Würzburg, Physikalisches Institut and Würzburg-Dresden Cluster of Excellence ct.qmat, Lehrstuhl für Technische Physik, Am Hubland, 97074 Würzburg, Germany}
\author{Alexander W. Holleitner}
\affiliation{Walter Schottky Institut and TUM School of Natural Science, Technische Universit{\" a}t M{\" u}nchen, Am Coulombwall 4, 85748 Garching, Germany}
\affiliation{Munich Center for Quantum Science and Technology (MCQST), Schellingstr. 4, 80799 Munich, Germany}
\author{Jonathan J. Finley}
\email{finley@wsi.tum.de}
\affiliation{Walter Schottky Institut and TUM School of Natural Science, Technische Universit{\" a}t M{\" u}nchen, Am Coulombwall 4, 85748 Garching, Germany}
%\date{\today}

\begin{abstract}
    We report lasing of moiré trapped interlayer excitons (IXs) by integrating a pristine hBN-encapsulated MoSe$_2$/WSe$_2$ heterobilayer into a high-$Q$
    ($>10^4$) nanophotonic cavity.
    We control the cavity-IX detuning using a magnetic field and measure their dipolar coupling strength to be $78 \pm 4\ \mathrm{\mu eV}$, fully consistent with the 82 $\mathrm{\mu eV}$ predicted by theory.
    The emission from the cavity mode shows clear threshold-like behavior as the transition is tuned into resonance with the cavity.
    We observe a superlinear power dependence accompanied by a narrowing of the linewidth as the distinct features of lasing.
    The onset and prominence of these threshold-like behaviors are pronounced at resonance while weak off-resonance.
    Our results show that a lasing transition can be induced in interacting moiré IXs with macroscopic coherence extending over the length scale of the cavity mode.
    Such systems raise interesting perspectives for low-power switching and synaptic nanophotonic devices using two-dimensional materials.
\end{abstract}

\maketitle

\section*{Introduction}

Transition metal dichalcogenides (TMDs) are atomically thin semiconductors that attract broad attention due to their novel electronic, optical and photoelastic properties \cite{Manzeli2017,RevModPhys.90.021001,Tartakovskii2020}.
The photophysics of TMDs is dominated by excitonic phenomena with large binding energies \cite{PhysRevLett.120.057405,Goryca2019}.
A particularly rich field of research concerns the formation of homo- \cite{PhysRevLett.130.026901} and hetero- \cite{Rivera2018,Regan2022} twisted bilayers, for which the twist-angle dependent hybridization of frontier orbitals results in novel bandstructure and the formation of interlayer excitons (IXs) where electron and hole are primarily localized in adjacent monolayer materials.
When the TMD heterostructures are fully encapsulated by a homogeneous dielectric such as hBN, the exciton linewidths reduce close to the Fourier limit, and the optical emission energies become stabilized \cite{PhysRevX.7.021026,Wierzbowski2017,Raja2019}.
The ability to tailor the local energy landscape experienced by excitons provides pathways to single photon emission.
Specific examples include the atomic vacancy \cite{Klein2019,Mitterreiter2021} or the moiré superlattice potential induced in twisted heterobilayer TMDs \cite{Seyler2019,doi:10.1126/sciadv.aba8526,Jiang2021}.
Such localized excitons have been shown to have quantum-dot-like electro-optical properties such as a discrete emission spectrum and on-demand single photon generation.
In particular, the moiré potential is spatially periodic in the superlattice, and the trapped IXs inherit the spin-valley contrasting properties of the heterobilayer components \cite{Seyler2019,doi:10.1126/sciadv.aba8526,Jiang2021}.
Such unique properties make them highly interesting for exploring correlated phases of electrons and coherent spin-photon interfaces, e.g., through the coupling to a local optical field in a nanocavity.

Nanocavity-TMD lasing has been obtained by coupling free excitons in monolayer TMD \cite{Wu2015,Ye2015,Li2017} or IXs in heterobilayer TMDs \cite{Paik2019,doi:10.1126/sciadv.aav4506,doi:10.1002/lpor.202000271,2302.01266} to prefabricated cavities using the pick-and-place methods.
These excitons have a continuous and spectrally broad gain spectrum, and their lasing in the cavity is described using coupled rate equation models describing the excitons and photons.
In contrast, single quantum emitter with discrete gain spectrum, such as quantum dot, enables the nanocavity laser at the ultimate quantum limit and interesting properties such as the on-demand generation of nonclassical states of light \cite{Carmichael2003,Nomura2010,Ritter:10,doi:10.1002/lpor.201000039,Lichtmannecker2017}.
Moiré trapped IXs also have discrete energy spectrum for single-photon emission \cite{doi:10.1126/sciadv.aba8526} and, moreover, inherit features of the van der Waals heterostructure such as strong interparticle interactions that can be used to build synaptic nanophotonic devices \cite{10.1002/adfm.202005443}.
However, in previous pick-and-place cavities, the excitonic energy landscape is strongly disordered by the complex dielectric screening of the substrate \cite{PhysRevX.7.021026,Wierzbowski2017,Raja2019}.
The local exciton trapping potential becomes dominated by the dielectric environment, rather not by the frontier orbital hybridization.
Therefore, cavity quantum electrodynamics (QED) studies of native moiré trapped IXs with the discrete energy spectrum remain a challenge.
Recently, we demonstrated how pristine two-dimensional (2D) material heterostructures can be integrated into high-$Q$ nanophotonic cavities, to realize a cavity QED platform in which light-matter interactions can be controlled \cite{PhysRevLett.128.237403,PhysRevLett.130.126901}.

\begin{figure*}
    \includegraphics[width=\linewidth]{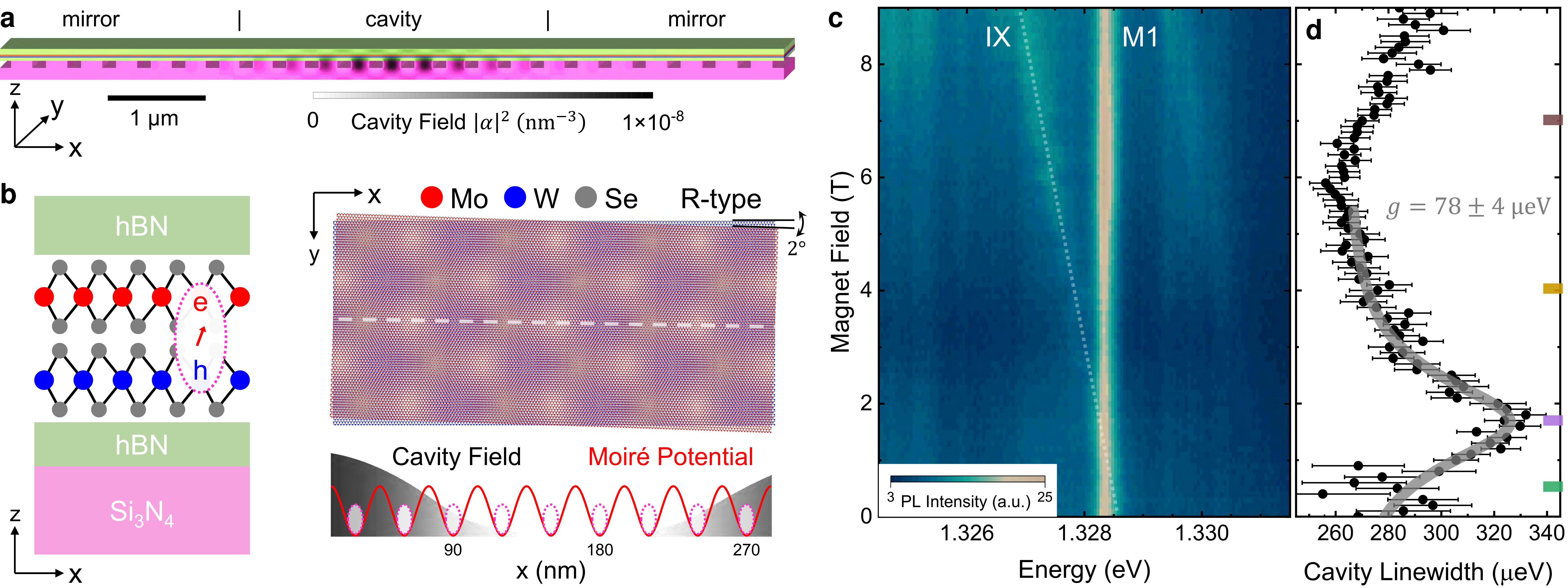}
    \caption{\label{f1}
        Cavity-trapped IX coupling at low excitation power.
        (a) Structure of the nanobeam cavity and the electric field profile $\vert\alpha\vert^2$ of cavity mode M1.
        (b) Schematic of the R-type MoSe$_2$/WSe$_2$ heterobilayer embedded in the cavity and their moiré superlattice.
        The black pattern denotes a small period of cavity electric field around the nanobeam center.
        (c) Magneto PL spectra recorded at a low excitation power of 88 nW.
        a.u., arbitrary units.
        (d) The detuning dependence of cavity linewidth at the right
        panel gives the exciton-photon
        coupling strength of $78 \pm 4\ \mathrm{\mu eV}$.
    }
\end{figure*}

Here, we use this platform and report the observation of lasing from moiré trapped IXs coupled to a high-quality factor ($Q$) ($>10^4$) mode of nanobeam photonic crystal cavity.
The nanocavity is embedded with a fully encapsulated MoSe$_2$/WSe$_2$ heterobilayer that is prepared to be R-type with a twist angle of 2$^\circ$, identified by the Zeeman splitting and negative circular polarization of IX \cite{Seyler2019,doi:10.1126/sciadv.aba8526}.
We apply a magnetic field in the Faraday geometry to tune a single IX line relative to the cavity mode \cite{Seyler2019}.
The photoluminescence (PL) emission from the cavity mode shows a linewidth broadening at resonance under the weak excitation, indicative of the cavity-trapped IX coupling.
We extract a coupling strength of $78 \pm 4\ \mathrm{\mu eV}$ from the experiment, fully consistent with the $82$ $\mathrm{\mu eV}$ predicted by the Tavis-Cummings model in theory.
Moreover, when the IX is close to resonance with the cavity mode (detuning of $<230\ \mathrm{\mu eV}$), a super linear intensity-pump power ($L$-$L$) relationship is observed, accompanied by linewidth narrowing as the distinct feature of a transition to lasing \cite{Paik2019,doi:10.1126/sciadv.aav4506,doi:10.1002/lpor.202000271}.
This threshold-like feature rapidly becomes weaker off-resonance (detuning of $>460\ \mathrm{\mu eV}$), indicating that a transition to lasing is observed at resonance because other features such as experimental fluctuations are not detuning-dependent \cite{Nomura2010,PhysRevB.81.033309}.
Our results show that moiré trapped IXs have sufficient oscillator strength to achieve lasing in high-$Q$ cavity QED devices, thereby exhibiting macroscopic coherence that extends over the length scale of the cavity mode.

\section*{Results}

The structure of our nanobeam cavities is presented schematically in Fig. \ref{f1}(a).
The photonic crystal lattices at either side of the cavity region have a constant periodicity and function as mirrors, to confine photons at the center where the periodicity is locally chirped over several unit cells using a Gaussian function \cite{PhysRevLett.128.237403}.
The calculated electric field profile of cavity mode M1 shows a mode volume of
1.2$\left(\lambda/n\right)^3$, where n is the effective index of the guided mode, as depicted schematically in Fig. \ref{f1}(a).
The hBN-encapsulated MoSe$_2$/WSe$_2$ heterostructure was prepared using mechanical exfoliation and viscoelastic dry transfer method \cite{Watson_2021}.
The heterobilayer is R-type with a twist angle of $\sim 2^\circ$, resulting in a moiré superlattice with lattice periodicity of $\sim 30$ nm, as depicted in Fig. \ref{f1}(b).
This moiré superlattice provides a smooth confinement potential to trap the IXs involving electrons in the conduction band of MoSe$_2$ and holes in the valence band of WSe$_2$.
Such moiré trapped IXs inherit the valley contrasting properties of the  eterobilayer such as the circular polarization and Zeeman shift \cite{Seyler2019}.
The nanobeam width (540 nm) is much larger than the moiré trap size and will not, therefore, affect the trapped IX.
Detailed fabrication procedures of the investigated samples are presented in Sec. \ref{secs1a} in the Supplementary Materials.

We performed magneto PL spectroscopy at low temperature ($T=10$ K) using a confocal micro-PL setup with the out-of-plane magnetic field tuned from 0 to 9 T.
The sample was excited using a continuous-wave (CW) laser at 640 nm focused to a spot size of $\sim 1\ \mathrm{\mu m}$.
In Fig. \ref{f1}(c), we present typical PL spectra recorded at low excitation power (88 nW), including the emission from the cavity mode M1 and one IX line.
The high-$Q$ factor of mode M1 ($Q=12500$) was confirmed using high-resolution
spectroscopy and coherence time measurement (for details, see Sec. \ref{secs1d} in the Supplementary Materials).
The IX exhibits a linear Zeeman shift of $197\ \mathrm{\mu eV/T}$ corresponding to an effective g factor $\lvert g_\mathrm{eff}\rvert=6.8$.
This value of $g_\mathrm{eff}$ is a signature of a MoSe$_2$/WSe$_2$ heterobilayer with the twist angle of 2$^\circ$ \cite{Seyler2019,Jiang2021,Hsu2018,PhysRevB.101.235408}.
Moreover, all trapped IXs are found to have a similar $g_\mathrm{eff}$ and exhibit a negative circular polarization degree (for details, see Sec. \ref{secs2} in the Supplementary Materials).
These uniform-valley contrasting properties are distinct features of the moiré potential in contrast to other randomly formed extrinsic traps \cite{Seyler2019,doi:10.1021/acs.nanolett.1c01215,Mahdikhanysarvejahany2022}.

At low excitation power, the coherent exciton-photon energy exchange dominates the system as described by the Jaynes-Cummings and Tavis-Cummings models, depending on the number of emitters coupled to the cavity mode.
The linewidth of the cavity mode is then \cite{PhysRevLett.128.237403,PhysRevB.82.045306,10.1093/oso/9780198768609.003.0008}
\begin{eqnarray}
    \label{width}
    \gamma_-=\gamma_C+\frac{g^2\Delta\gamma}{\Delta\omega^2+\frac{1}{4}\Delta\gamma^2}
\end{eqnarray}
where $\gamma_-$ is the measured cavity linewidth broadened by the coupling, $\gamma_C$ is the bare cavity linewidth, $\Delta\omega$ ($\Delta\gamma$) is the energy detuning (linewidth difference) between the bare cavity mode and bare trapped IX, and $g$ is the exciton-photon coupling strength, i.e., the
rate at which energy is coherently exchanged between the coupled
systems.
In our experiments, we primarily control the detuning $\Delta\omega=\frac{1}{2}g_\mathrm{eff}\mu_\mathrm{B}\left(B-B_0\right)$ using the magnetic field $B$, where $\mu_\mathrm{B}$ is the Bohr magneton and $B_0$ is the magnetic field at resonance.
The variation of $\gamma_C$ and $\Delta\gamma$ in the measurement is small compared to the variation of $\Delta\omega$, and, hence, we take these parameters to be constant.
We then use Eq. \ref{width} to fit the measured $\gamma_-$ as presented by the gray peak in Fig. \ref{f1}(d) and determine the coupling strength to be $g=78 \pm 4\ \mathrm{\mu eV}$, a value comparable to traditional discrete quantum emitters, such as III-V quantum dots \cite{RevModPhys.87.347,PhysRevLett.122.087401}.
From the fit, we also retrieve the trapped IX linewidth to be $640 \pm 40\ \mathrm{\mu eV}$, fully consistent with the value directly measured when the IX is strongly detuned from M1 ($B>6\ \mathrm{T}$).
These values reveal that, if the moiré IX has a narrower linewidth such as $\sim 100\ \mathrm{\mu eV}$ at ultralow temperatures below 1 K \cite{PhysRevLett.126.047401,doi:10.1021/acsnano.0c08981,Mahdikhanysarvejahany2021} or if the cavity couples to more IX lines, then the exciton-photon interaction will enter the strong coupling regime of the light-matter interaction \cite{10.1093/oso/9780198768609.003.0008}.
For another cavity mode that couples to multiple IX lines, we observe nonlinear magneto energy shifts in the PL spectra, which are typical features of polaritons in the strong coupling regime (see Sec. \ref{secs3b} in the Supplementary Materials for details).

\begin{figure}
    \includegraphics[width=\linewidth]{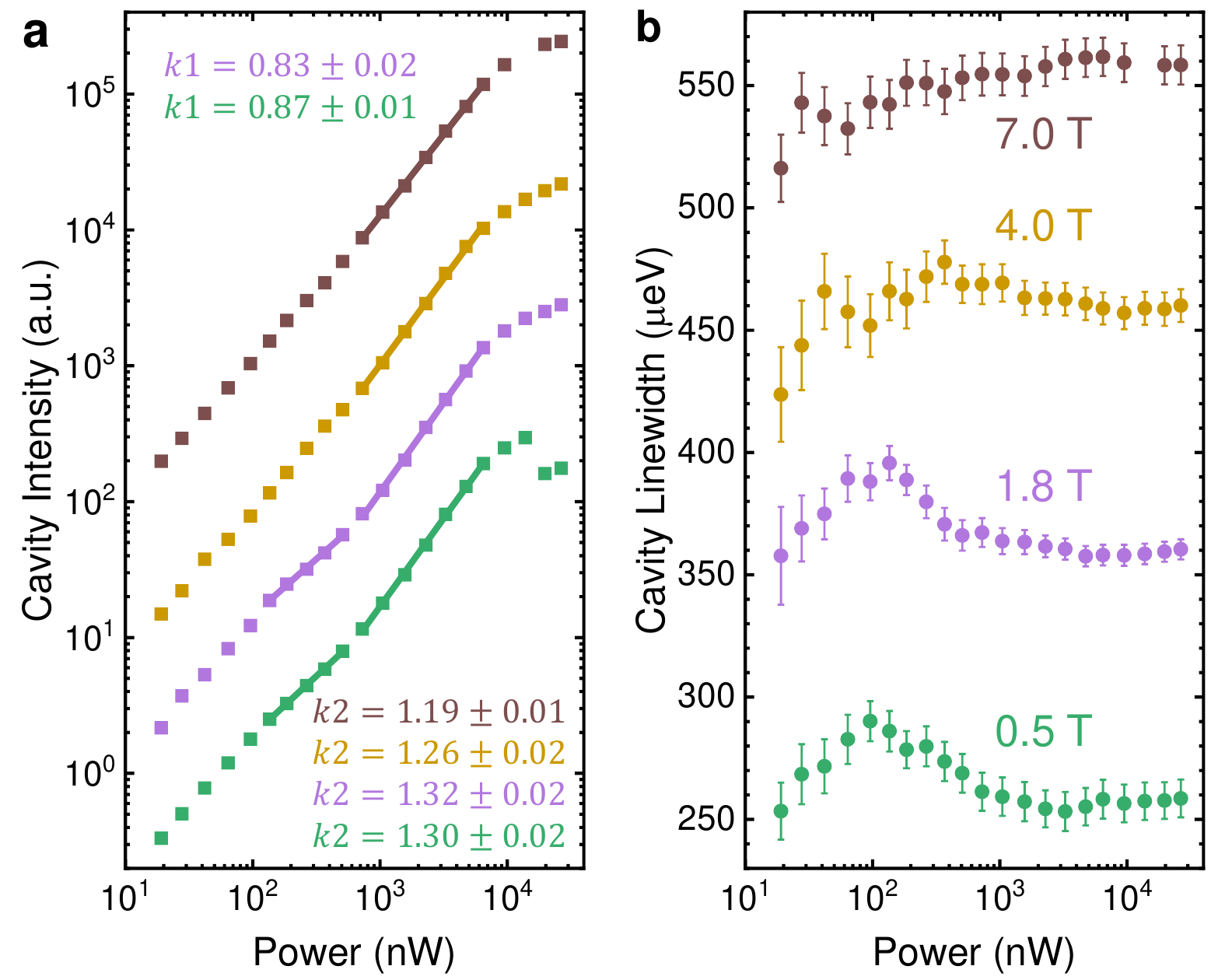}
    \caption{\label{f2}
    Lasing of trapped IX at high excitation power and the detuning dependence.
    Power-dependent (a) intensity $L$ and (b) linewidth of the cavity mode measured at the magnetic field 0.5, 1.8, 4.0, and 7.0 T, corresponding to the detuning $\Delta\omega$ of $-0.23$, $0.02$, $0.46$ and $1.05$ meV, respectively.
    As $\Delta\omega$ increases, the lasing features including the nonlinearity of the exponent on the $L$-$L$ curve $k$ and the pump-induced narrowing of linewidth evolve from pronounced to weak and lastly suppressed.
    }
\end{figure}

We continue to explore the power-dependent emission for different cavity-trapped IX detunings as controlled by the magnetic field.
In Fig. \ref{f2}, we present typical results recorded for $B =$ 0.5, 1.8, 4.0, and 7.0 T, corresponding to detunings of $\Delta\omega=$ $-0.23$, 0.02, 0.46, and 1.05 meV.
As presented in Fig. \ref{f2}(a), the intensity of the cavity mode closer to resonance ($\Delta\omega=0.02$ meV) exhibits a clear superlinear threshold-like
behavior for which the exponent of the $L$-$L$ curve is linear $k_0=1.04\pm0.04$ at low excitation power of $<10^{2}$ nW, and then sublinear $k_1=0.83\pm0.02$ and increases again to $k_2= 1.32\pm0.02$ at a threshold.
Meanwhile, as presented in Fig. \ref{f2}(b), the linewidth narrowing of the cavity mode corresponding to the increasing coherence of the emitted optical field is also observed, and the threshold for linewidth narrowing $10^{2}$ nW is the same value for the sublinearity.
These observations are both distinct features for a nanocavity laser hosting excitons with discrete energy spectrum \cite{Nomura2010,Ritter:10,doi:10.1002/lpor.201000039,Lichtmannecker2017}.
We note that at resonance the cavity linewidth narrows from 296 to 258 $\mathrm{\mu eV}$.
These raw values are broadened by the $160$-$\mathrm{\mu eV}$ resolution limit of the used spectrometer and were checked using another high-resolution spectrometer and measurements of the coherence time of the emitted field (for details see Sec. \ref{secs3a} in the Supplementary Materials).
After deconvolution for the resolution, the observed linewidth narrowing corresponds to 136 to 98 $\mathrm{\mu eV}$ (28$\%$) as the system is pumped through threshold.
Moreover, these two characteristic lasing features (threshold and linewidth narrowing) are only found to be pronounced at or near resonance ($\Delta\omega=-0.23$ and $0.02$ meV).
They are much weaker for small detunings ($\Delta\omega=0.46$ meV) and entirely
suppressed when strongly detuned ($\Delta\omega=1.05$ meV).
The clear detuning dependencies obtained from Fig. \ref{f2} further strengthen the identification of lasing from the cavity-trapped IX coupling, because other features such as slow experimental fluctuations are not detuning-dependent
and the results presented were found to be reproducible and only sensitive to $\Delta\omega$.

\begin{figure}
    \includegraphics[width=\linewidth]{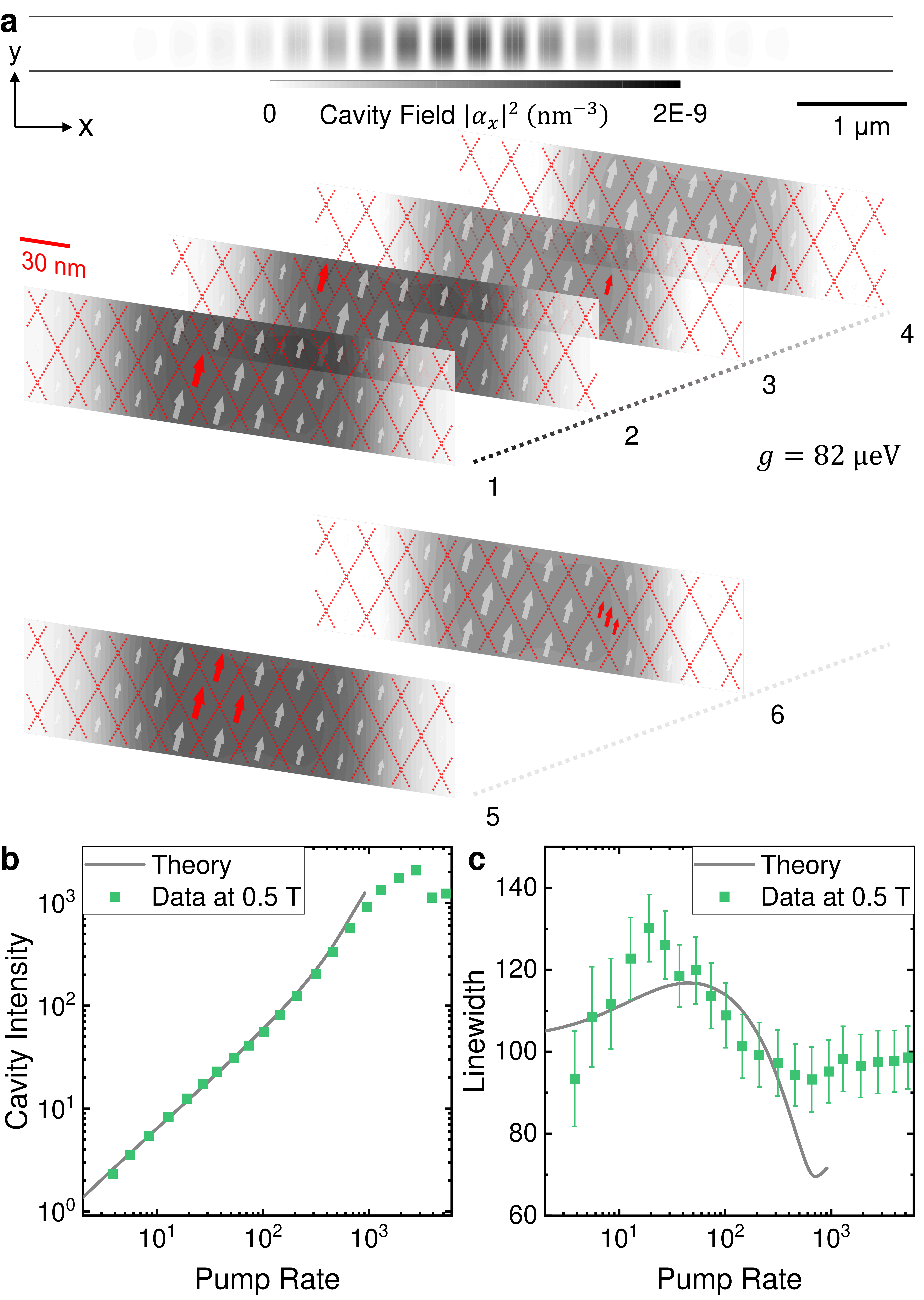}
    \caption{\label{f3}
    Theoretical calculation of the cavity-trapped IX coupling and lasing.
    (a) Panels 1 to 4 are schematic representations of different snapshots illustrating the exciton-photon interaction at resonance.
    Panels 5 and 6 are snapshots for the situation when the coupling is detuned by the exciton-exciton interactions, leading to an incoherent feeding of cavity photons.
    (b) Intensity and (c) linewidth of the cavity mode calculated by master equation theory showing that it is in very good accordance with the experimental data.
    }
\end{figure}

Our experimental observations are well reproduced by numerical modeling.
A single IX line is a single-photon emitter \cite{doi:10.1126/sciadv.aba8526}, indicating excitation of a single trapped IX, as schematically depicted in Fig. \ref{f3}(a) (top).
This means that, for the exciton-photon coupling at resonance, only a single trapped IX is excited simultaneously, and this IX can occupy any one of the moiré sites within the spatial area overlapping with the cavity mode.
This situation is analogous to the Tavis-Cummings model of cavity QED with multiple emitters.
In contrast, when multiple excitations of IXs exist within one moiré site or several neighboring sites, the exciton-exciton interaction will energetically shift and broaden the emission \cite{Kremser2020,Mahdikhanysarvejahany2021,PhysRevX.11.031033,Sun2022}, thereby detuning the exciton-photon coupling, as schematically depicted in Fig. \ref{f3}(a) (bottom).
This leads to an incoherent feeding of photons into the cavity mode via mechanisms similar to the case of nanocavity lasers having a few quantum dots as gain medium \cite{Carmichael2003,Nomura2010,Ritter:10,doi:10.1002/lpor.201000039,Lichtmannecker2017}.
To describe the situation in our experiment, we first calculate the coherent
exciton-photon coupling strength using the Tavis-Cummings model.
The quantized electric field of the cavity mode is
\begin{eqnarray}
    \mathbf{E}\left(\mathbf{r}\right)=i\left(\frac{\hbar\omega_{a}}{2\epsilon_r\epsilon_0}\right)^{{1}/{2}}\lbrack a\bm{\alpha}\left(\mathbf{r}\right)-a^{+}\bm{\alpha^\ast}\left(\mathbf{r}\right) \rbrack \nonumber
\end{eqnarray}
where $a^{+}$ and $a$ are the ladder operators of photons corresponding to the cavity and $\bm{\alpha}\left(\mathbf{r}\right)$ is the normalized amplitude of the cavity field at position $\mathbf{r}$.
$\hbar\omega_a$ is the cavity energy, and $\epsilon_r$ ($\epsilon_0$) is the relative (vacuum) permittivity.
$\bm{\alpha}\left(\mathbf{r}\right)$ is extracted from 3D finite difference time
domain (FDTD) simulations, and the results are presented in Figs. \ref{f1}(a) and \ref{f3}(a).
These simulations account very well for the observed cavity mode energy and dependence on the width of the nanocavity beam \cite{PhysRevLett.128.237403}.
Moreover, we note that the trapped IX is confined within a single lattice site with the size of $\sim 30$ nm.
This is much smaller than the photon wavelength 933 nm, and, therefore, we expect the dipole approximation to be entirely valid in this case for localized excitons.
Because the optical dipole of IX is in-plane \cite{PhysRevB.105.035417}, the coupling between the cavity mode and the different trapped IX sites is then described by the Tavis-Cummings model
\begin{eqnarray}
    \sum\limits_i\hbar g_i\left(\sigma_{Xi,Gi}a+\sigma_{Gi,Xi}a^{+}\right),\ \hbar g_i=\left(\frac{\hbar\omega_{a}}{2\epsilon_r\epsilon_0}\right)^{{1}/{2}}d\alpha_x\left(\mathbf{r}_i\right) \nonumber
\end{eqnarray}
where $g_i$ is the coupling strengh of the IX at position $\mathbf{r}_i$ and $d$ is the optical dipole moment of IX.
$\sigma_{Xi,Gi}$ and $\sigma_{Gi,Xi}$ are the Dirac operator for the IX $i$ described by the excited level $X_i$ and ground level $G_i$.
The collective coupling strength g is calculated in the weak excitation limit, i.e., only one IX is at the excited level as denoted by the red spins in Fig. \ref{f3}(a).
In this case, the coupling is equal to that between one photon in the cavity mode and a collective emitter.
The excited level of the collective emitter is $X=\sum_ig_iX_i/g$, where the trapped IXs within the cavity mode volume have macroscopic coherent phases.
The collective coupling strength is
\begin{eqnarray}
    \hbar g=\hbar\left(\sum\limits_i g_i^2\right)^{{1}/{2}}=\left(\frac{\hbar\omega_{a}}{2\epsilon_r\epsilon_0}\right)^{{1}/{2}}d\alpha_{\mathrm{sum}} \nonumber
\end{eqnarray}
where
\begin{eqnarray}
    \alpha_{\mathrm{sum}}=\left(\sum\limits_i\vert\alpha_x\left(\mathbf{r}_i\right)\vert^2\right)^{{1}/{2}} \nonumber
\end{eqnarray}
represents the integration over all moiré superlattice sites that lie spatially within the cavity mode volume.
On the basis of the cavity field profile and the size of the moiré site, as shown in Fig. \ref{f3}(a), we numerically integrate and obtain a value $\alpha_{\mathrm{sum}}=9.1\times10^{-4}$ $\mathrm{nm^{-1.5}}$.
The optical dipole moment $d$ can be calculated from the IX radiative lifetime $\tau_\mathrm{rad}$ by \cite{KARRAI2003311}
\begin{eqnarray}
    \hbar \gamma_\mathrm{rad}=\frac{8\pi^2n}{3\lambda^3}\frac{d^2}{\hbar\epsilon_0} \nonumber
\end{eqnarray}
where $\gamma_\mathrm{rad}=1/\tau_\mathrm{rad}$ is the radiative decay rate, $n=2.0$ is the environment (hBN) refractive index, and $\lambda=933$ nm is the emission wavelength.
For the R-type heterobilayer in our cavity, the radiative lifetime of the moiré trapped IX is 200 ns \cite{Mahdikhanysarvejahany2021}, corresponding to the optical dipole moment $d=52\ \mathrm{e\cdot pm}$.
This is significantly smaller than a quantum dot ($\sim 0.6\ \mathrm{e\cdot nm}$), consistent with the much weaker emission of the trapped IX.
In addition, this dipole moment $52\ \mathrm{e\cdot pm}$ is also much smaller than the magnetic confinement length $\sqrt{\hbar/eB}$ ($\sim 9$ nm at 9 T) \cite{PhysRevLett.103.127401,PhysRevLett.122.087401}, and it is therefore unaffected by the magnetic field.
We then use $\omega_a=1.33$ eV and $\epsilon_r=n^2=4.0$ to estimate the coupling strength, obtaining a value of $\hbar g=82$ $\mathrm{\mu eV}$.
This result is fully consistent with the $78 \pm 4\ \mathrm{\mu eV}$ from the experimental data in Fig. \ref{f1}.

The lasing of discrete localized excitons arises from both the coherent exciton-photon coupling and the incoherent feeding of cavity photons shown in Fig. \ref{f3}(a), and the conventional rate equation for free excitons is insufficient to describe such lasing at the quantum limit \cite{doi:10.1002/lpor.201000039}.
Hereby, we use the master equation theory to simulate the pump-dependent transition to lasing oscillation in the system, following the approach made by Nomura et al. \cite{Nomura2010}.
Specific details can be found in Sec. \ref{secs4} in the Supplementary Materials.
Typical results calculated for the case of 200 $\mathrm{\mu eV}$ detuning ($B=0.5$ T) are presented in Fig. \ref{f3}(b) and (c).
The pump-dependent cavity intensity in Fig. \ref{f3}(b) reproduces our experimental data well.
The simulated cavity linewidth in Fig. \ref{f3}(c) qualitatively describes our experimental observation but with minor differences, which, we believe, originate from experimental causes that have not been considered in the calculation.
The spectrometer used has a comparatively short focal length (0.3 m), which introduces a spectral broadening of $160$ $\mathrm{\mu eV}$ as discussed in the context of Fig. \ref{f2}(b).
This broadening is deconvoluted from the experimental data in Fig. \ref{f3}(c).
In addition, the coupling between intensity and phase noise \cite{Wu2015,doi:10.1002/lpor.201000039,doi:10.1063/1.106693,doi:10.1063/1.105443} widely exists in nanocavity lasers and further broadens the linewidth at high excitation power.
This is not included in the calculation.
Despite these minor differences, the calculations in Fig. \ref{f3} provide a good description of the experimental data, thereby strongly supporting our conclusion of the coupling and lasing in the cavity-trapped IX system.

\begin{figure}
    \includegraphics[width=\linewidth]{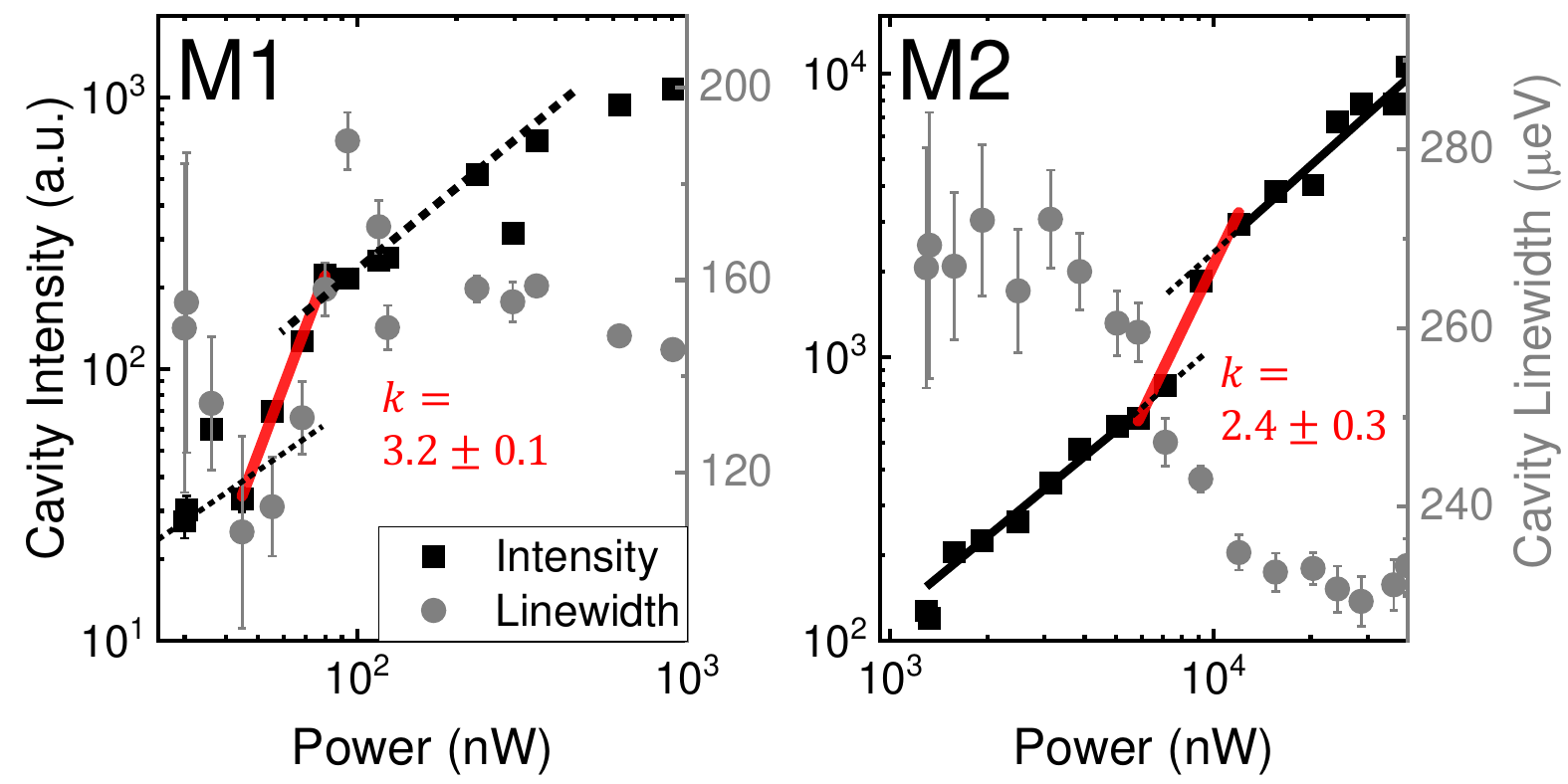}
    \caption{\label{f4}
    Lasing under green laser excitation.
    Power-dependent intensity $L$ and linewidth of two cavity modes, when excited by a green laser.
    In this case, the lasing features are more pronounced.
    }
\end{figure}

Furthermore, we also found that specific fingerprints of lasing are related to the excitation conditions.
Here, we performed PL spectroscopy without a magnetic field but excited using a 532-nm green CW laser instead of the red pump laser used earlier.
Typical data are presented in Fig. \ref{f4}, showing the lasing features of mode M1 as well as another mode M2.
The superlinear exponent of the $L$-$L$ curve increases to 3.2 $\pm$ 0.1 for the mode M1 when subject to green laser excitation, much larger than the value of 1.3 observed using red laser (Fig. \ref{f2}).
Similarly, mode M2 also exhibits a larger superlinear exponent of 2.4 $\pm$ 0.3.
Meanwhile, the narrowing linewidth of the cavity mode is observed to occur simultaneously with the superlinearity of the $L$-$L$ curve.
In general, the lasing features when subject to green laser excitation are more pronounced.
We explain these differences by the different background emission under different excitation conditions.
Compared to the red laser, the green laser is expected to generate more dephasing effects because the thermal energy generated per absorbed pump photon is much larger ($\sim 1$ eV for 532 nm, cf. 0.6 eV for 640 nm).
This is likely to lead to the formation of an exciton gas with higher steady-state temperature, and IXs are less easily trapped in the moiré potential.
This, in turn, leads to stronger feeding of cavity photons, thereby enhancing the nonlinearity (Fig. \ref{sf11}).
The dephasing effects also account well for the re-broadening of cavity linewidth toward the strongest pumping.
For mode M1, the re-broadening under the green laser excitation (Fig. \ref{f4}) is more pronounced compared to that under the red laser excitation (Fig. \ref{f2}).
As discussed in the context of Fig. \ref{f3}(c), this re-broadening arises from the coupling between intensity and phase noises \cite{doi:10.1063/1.106693,doi:10.1063/1.105443}.
As such, the enhanced re-broadening observed for M1 is consistent to the larger dephasing effects discussed above.
In contrast, the re-broadening is not observed for mode M2 in Fig. \ref{f4}.
Mode M2 is at 1.36 eV, strongly detuned to the delocalized IXs.
As such, fewer delocalized IXs lead to lower phase noise theoretically, consistent to the suppression of re-broadening for mode M2 in the experiment.

\section*{Discussion}

In summary, we report the observation of lasing for moiré trapped IXs by magnetically tuning the IXs in a high-$Q$ nanobeam cavity embedded with hBN-encapsulated MoSe$_2$/WSe$_2$ heterobilayer.
We measure the coupling strength of a single IX line to the high-$Q$ cavity
mode to be 78 $\mathrm{\mu eV}$, a value comparable to traditional emitters
\cite{RevModPhys.87.347,PhysRevLett.122.087401} and indicating the potential of strong coupling regime with ultralow temperatures or multiple IX lines.
The stimulated lasing driven by the cavity electric field enables the macroscopic coherence of moiré trapped IXs that extends over the length scale of the cavity mode.
Moreover, the homogeneous environment in our cavity also opens the way to apply more quantum dot-like emitters in 2D semiconductors, e.g., excitons trapped by the atomic vacancies \cite{Klein2019,Mitterreiter2021} and external electric field \cite{doi:10.1021/acs.nanolett.1c01215} into cavity QED devices.

Because the proximal interaction between different (magnetic, electric) degrees of freedom lies at the heart of the physics of 2D materials, such cavity QED can be used to modify the collective state of the system.
Compared to free excitons in 2D materials, the moiré IXs enable the nanocavity laser at the ultimate quantum limit and nonclassical properties \cite{Carmichael2003,Nomura2010,Ritter:10,doi:10.1002/lpor.201000039,Lichtmannecker2017}.
The discrete feature of moiré IX also allows the in situ control of the cavity-exciton detuning.
Thereby, we switched the system between nonlinear lasing regime at resonance and trivial linear regime off-resonance as shown in this work.
Such in situ control is difficult to achieve for the conventional laser with continuous gain spectrum.
Moreover, moiré IXs could form a dipolar exciton gas with strong interactions and thus have the potential for dipolar-mediated interaction between different lasing systems and optical competitions.

Compared to traditional discrete emitters such as quantum dot, the moiré IXs extend the physics and applications of the nanophotonic system.
For example, the integration of phase change or ferroelectric 2D materials adjacent to heterobilayers in the cavity may open routes toward optical devices exhibiting synaptic behavior and continuous actuation functions \cite{10.1063/1.5108562}, and the actuation power levels of such devices can potentially be engineered for low-power switching \cite{Zhu2022} with interesting potential for network architectures.
Meanwhile, the moiré excitons also improve the controllability of the coupling, because their opto-electronic properties and spatial distribution are determined by the hybridization of frontier orbitals, which can be controlled through the twist angle of the heterobilayer.
This paves the way to solve the long-standing problem of the random emission energy and the random spatial position of traditional self-assembled quantum dots.
As such, we believe that more nontrivial light-matter interaction will be revealed as the study of 2D material nanophotonics increases in the future.

\section*{Acknowledgments}

We acknowledge the German Science Foundation (DFG) for financial support via grants FI 947/8-1, DI 2013/5-1, HO 5194/16-1, and SPP-2244, as well as the clusters of excellence MCQST (EXS-2111), e-conversion (EXS-2089), and ct.qmat (EXC 2147, project-id 39085490).
J.J.F. and A.W.H. acknowledge the state of Bavaria via the One Munich Strategy and Munich Quantum Valley.
J.J.F. and S.H. acknowledge the state of Bavaria via the Integrated Spin Systems for Quantum Sensors (IQSense).
C.Q. and V.V. acknowledge the Alexander v. Humboldt foundation for financial support in the framework of their fellowship program.

\section*{Data Availability}

All data needed to evaluate the conclusions in the paper are present in the paper and/or the Supplementary Materials.
The datasets generated during and analyzed during the current study are available from doi: 10.5061/dryad.f1vhhmh38.

\section*{Author Contributions}
C.Q. and J.J.F. conceived and designed the experiments.
C.Q. and P.S. prepared the sample.
C.Q., M.T., J.F., V.V., and J.B. performed the experiments.
C.Q. analyzed the data and modeled the exciton-photon coupling.
All authors discussed the results and wrote the manuscript.

\section*{Competing Interests}
The authors declare that they have no competing interests.


% \newpage
\clearpage

\section*{Supplementary Materials}
\setcounter{figure}{0}
\renewcommand{\thefigure}{S\arabic{figure}}
\setcounter{equation}{0}
\renewcommand{\theequation}{S\arabic{equation}}

% \tableofcontents

\section{\label{secs1} Methods}

\subsection{\label{secs1a} Sample Fabrication}

\begin{figure}
    \includegraphics[width=\linewidth]{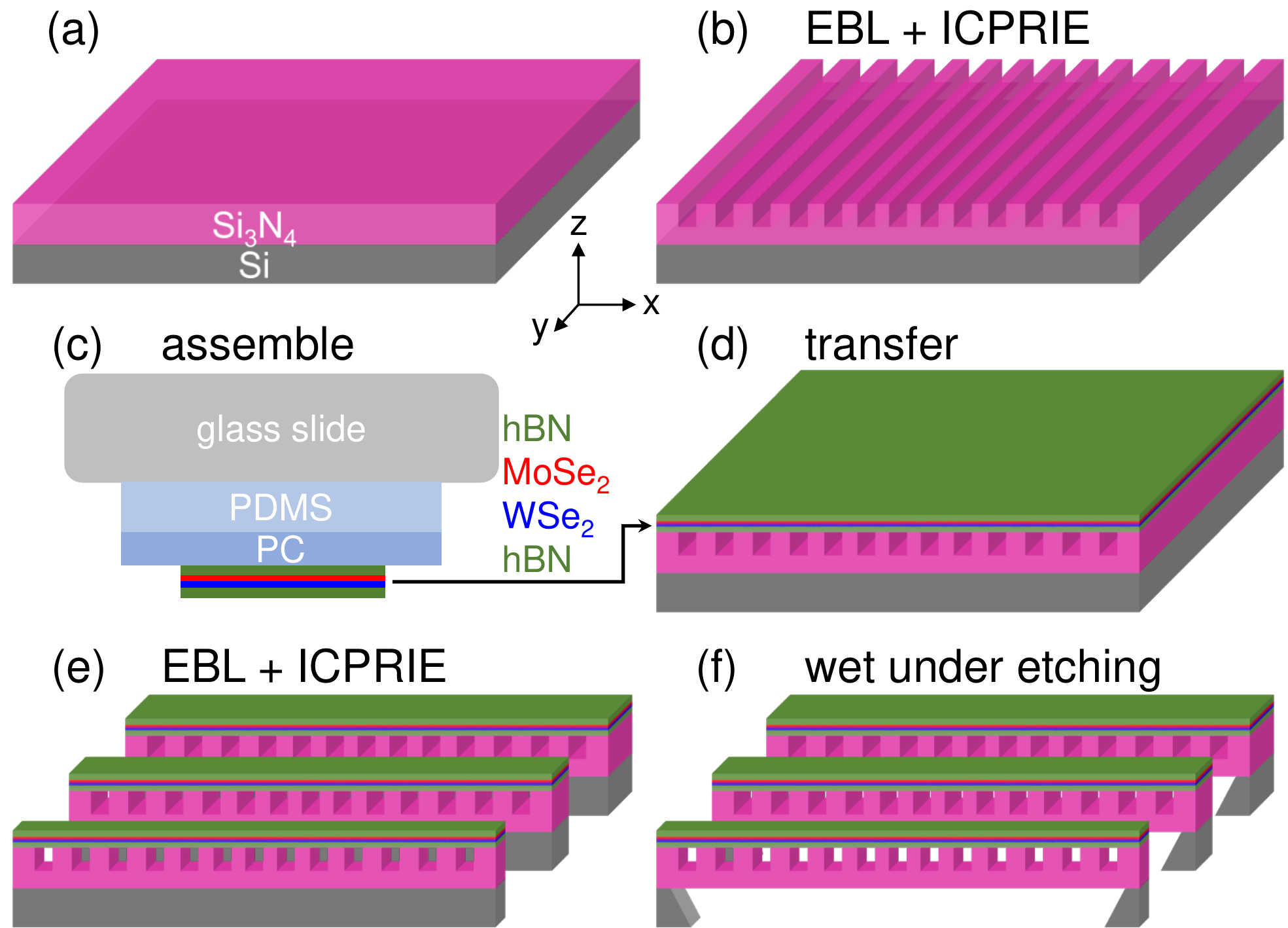}
    \caption{\label{sf1}
        Fabrication procedures.
        (a) Bare Si$_3$N$_4$/Si substrate.
        (b) Etch the periodic nanoscale trenches.
        (c) Assemble the heterostructure using a stamp and (d) transfer on cover the nanotrenches.
        (e) Divide the nanobeams.
        (f) Wet under etching to remove the bottom Si.
    }
\end{figure}

The fabrication procedures of our sample are schematically depicted in Fig.~\ref{sf1}.
Firstly, we prepare and clean the Si$_3$N$_4$/Si substrate, as shown in Fig.~\ref{sf1}(a).
The Si$_3$N$_4$ is grown by low pressure chemical vapor deposition (LPCVD) and has a thickness of 200 nm.
The Si at bottom is in 111 plane.
Then we use e-beam lithography (EBL) and inductively coupled plasma reactive ion etching (ICPRIE) to etch the periodic nanoscale trenches in Si$_3$N$_4$, as shown in Fig.~\ref{sf1}(b).
The nanotrenches have length $h_x=160\ \mathrm{nm}$ and depth $h_z=90\ \mathrm{nm}$, and follow a Gaussian distribution in spatial with the separation between them $a_i/a=1-A\cdot \mathrm{exp}(-i^2/(2\sigma^2))$.
$a=360$ nm is the lattice constant of photonic crystal.
$A=0.1$ and $\sigma=4$ define the chirped periodicity at center for a smoothly varying photonic confinement.
We assemble the hBN/MoSe$_2$/WSe$_2$/hBN heterostructure using the viscoelastic dry transfer method \cite{Watson_2021} and transfer the heterostructure on top of the nanoscale trenches, as shown in Fig.~\ref{sf1}(c)(d), respectively.
The top (bottom) hBN has the thickness 70 (40) nm.
Then we use the EBL and ICPRIE to divide the nanobeams, as shown in Fig.~\ref{sf1}(e).
These nanobeams has a varying width $d_y$ from 300 to 500 nm to tune the resonant energy of cavity mode.
Finally, we use 25$\%$ TMAH solution to under etch the bottom Si, as shown in Fig.~\ref{sf1}(f).
Detailed information about the recipe parameters and finite difference time domain simulation of cavity modes have been reported previously \cite{PhysRevLett.128.237403,PhysRevLett.130.126901}.

\begin{figure}
    \includegraphics[width=\linewidth]{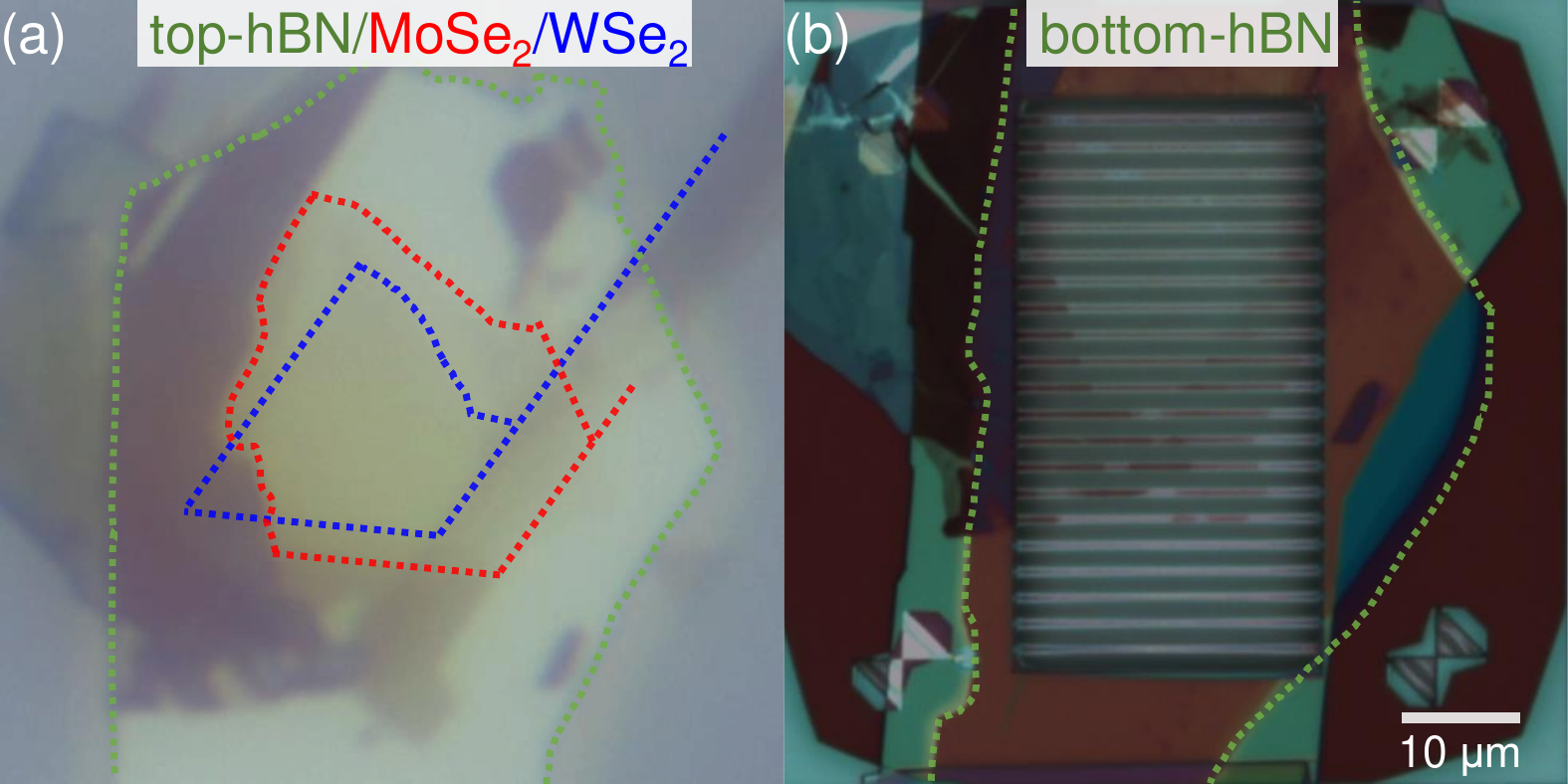}
    \caption{\label{sf2}
        Microscope images of samples.
        (a) Image of the stamp with top hBN, MoSe$_2$ and WSe$_2$.
        The monolayer regions of TMD flakes are aligned with the twist angle to either $\sim 0^\circ$ or $\sim 60^\circ$ as denoted in the figure.
        The bottom hBN has not been picked up at this moment.
        (b) Imange of the sample with the array of nanobeam cavities.
        The bottom hBN flake is denoted on the figure.
    }
\end{figure}

In Fig.~\ref{sf2}(a) we present the microscope image of the stamp (Fig.~\ref{sf1}(c)) when the top hBN, MoSe$_2$ and WSe$_2$ flakes are already picked up.
As denoted by the dashed lines, the monolayer regions of TMD flakes have some straight edges that reveal their lattice directions.
We align the monolayer MoSe$_2$ and WSe$_2$ with the parallel edges, which means the twist angle between them is either $\sim 0^\circ$ (R-type) or $\sim 60^\circ$ (H-type).
The R-type of our sample is identified by the Zeeman splitting and negative circular polarization \cite{Seyler2019,doi:10.1126/sciadv.aba8526}, as discussed later in Sec.~\ref{secs2}.
In Fig.~\ref{sf2}(b) we present the microscope image of the sample after fabrication (Fig.~\ref{sf1}(f)).
The array of bright nanobeams in the figure are the cavities.
Only some cavities are embedded with the heterobilayer TMDs, because the size of monolayer MoSe$_2$ and WSe$_2$ is $\sim 20\ \mathrm{\mu m}$ and cannot cover all the nanobeams.

Generally, moiré IXs require high-quality materials and homogeneous environment.
Therefore, the pick-and-place cavity used for free excitons, in which the non-encapsulated materials are transferred on top of a prefabricated nanocavity, cannot be applied to moiré IXs.
In contrast, owing to our cavity design, we mainly achieve three advantages compared to previous 2D-material cavities.
Firstly, we avoid perforating the 2D heterostructure, and thereby, achieve the pristine quality of moiré IXs.
Secondly, we achieve smooth nanoscale trenches etched in Si3N4, and thereby, improve the photonic Q-factors by one or two orders of magnitude compared to previous 2D-material cavities \cite{PhysRevLett.128.237403}.
Thirdly, in pick-and-place cavities the 2D materials are an attachment to the cavity and couple to only the evanescent field of cavity mode.
In this case, hBN encapsulation is usually not applied because it will decrease the exciton-evanescent-field overlap.
In contrast, in our cavity the hBN-encapsulated 2D heterostructure is one part of the cavity dielectric structure, and thereby, the 2D excitons are “really” integrated into the cavity with large overlap, rather not only couple to the evanescent field.
There three major advantages refer to the light, the matter, and the interaction, providing the basis of the lasing results reported in this work.

In addition, our cavity design also improves the long-term stability of the device.
It is well known that the long-term stability of 2D materials is not as good as bulk materials, and the optoelectronic properties are sensitive and could be affected by the disorder induced in experiments such as oxidation, doping, and reconstruction during the thermal cycling.
However, after encapsulation with hBN or other homogeneous materials, the sample is isolated from the environmental disorder, and the long-term stability of 2D materials has been reported to be greatly improved \cite{10.1021/acsnano.5b07677,doi:10.1021/acsami.2c12592}.
As a result, our sample was fabricated in February 2022, and the lasing datasets were measured in March 2022, August 2022, and November 2022, respectively, showing a good stability of at least eight months.

\subsection{\label{secs1b} Measurement Setup}

\begin{figure}
    \includegraphics[width=\linewidth]{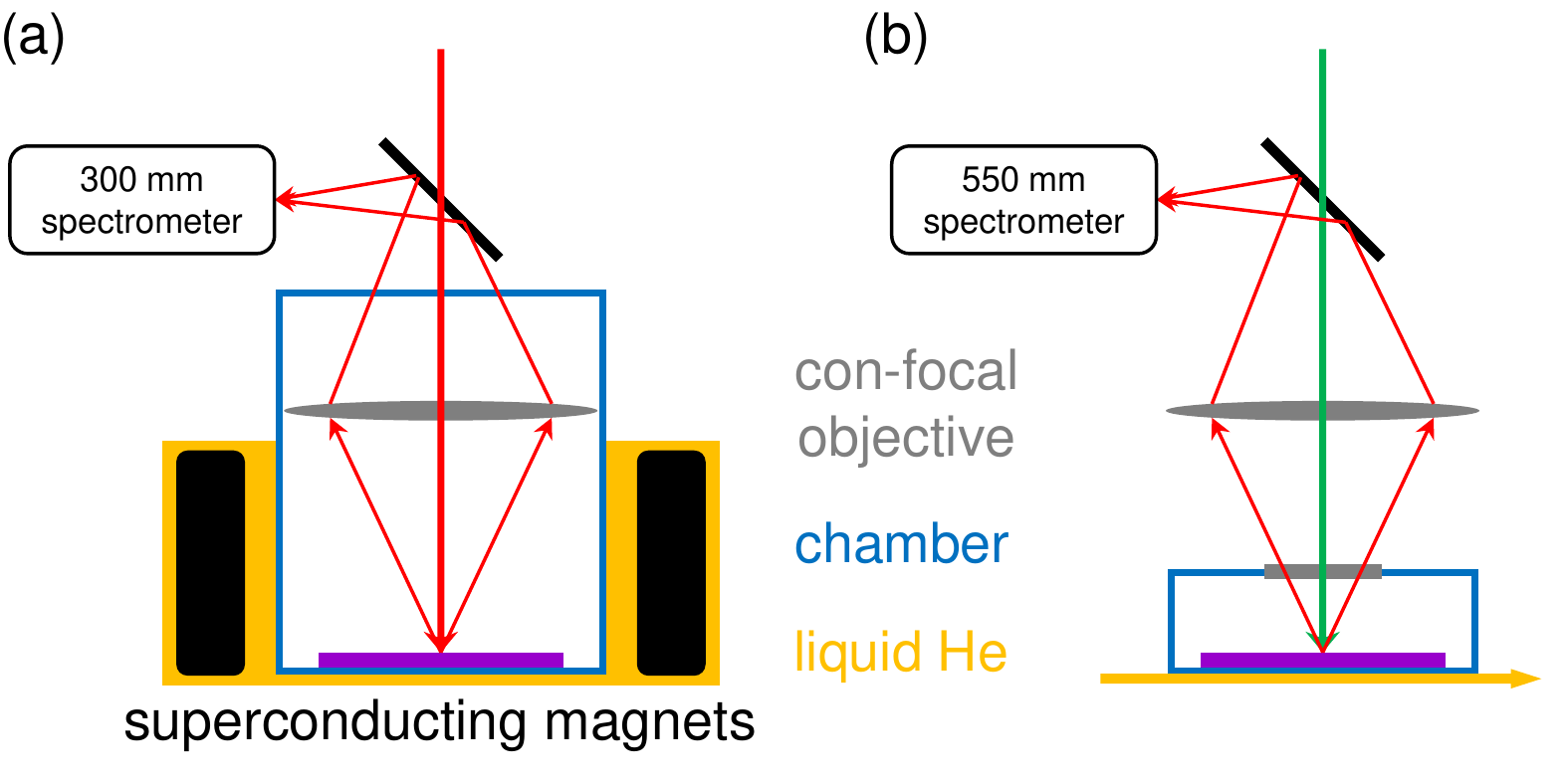}
    \caption{\label{sf3}
        Schematic of measurement setup.
        (a) Magneto micro-PL and (b) normal micro-PL setup.
        Different excitation laser and spectrometer are used in the two setups.
    }
\end{figure}

We use the confocal magneto micro-PL setup for the detuning dependent spectroscopy.
The schematic of this setup is shown in Fig.~\ref{sf3}(a).
The objective has the magnification of 100 and the NA of 0.81.
The sample is excited by the red cw-laser at 640 nm, and the spectra are collected by a matrix array Si CCD detector.
The spectrometer has the grating of 1200 grooves per mm whilst the focal length is 0.3 m.
Generally, a broadening is introduced in PL peaks by the spectrometer which reflects its resolution.
The shorter focal length results in larger instrumental broadening.
This broadening is $\sim160$ $\mathrm{\mu eV}$, extracted by the linewidth of singles from the excitation laser \cite{PhysRevLett.128.237403}.

In addition, we also implement the PL spectroscopy in a normal micro-PL setup without magnets, as shown in Fig.~\ref{sf3}(b).
This setup has a green cw-laser at 532 nm and the objective with the magnification of 100 and the NA of 0.75.
Moreover, the spectrometer has the same grating 1200 grooves per mm but the longer focal length 0.55 m, which provides a smaller instrumental broadening.
From our previous work \cite{2210.00150} we know that this broadening is approximately a Gaussian convolution with the width 50 pm.

\subsection{\label{secs1c} Fitting Methods}

\begin{figure}
    \includegraphics[width=\linewidth]{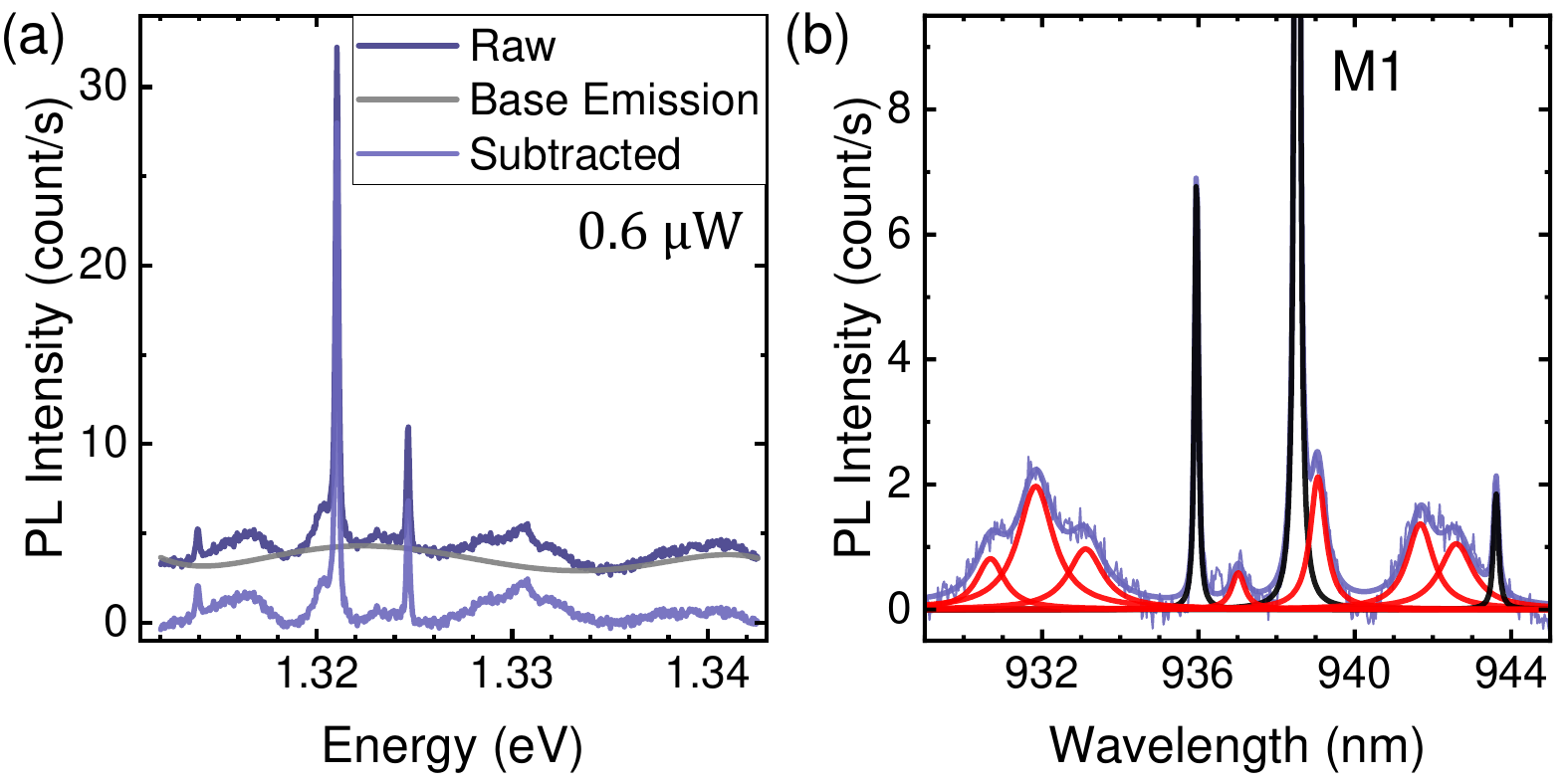}
    \caption{\label{sf4}
        Fitting examples.
        (a) Dark blue line is the raw data. Gray baseline is the emission from free and delocalized IXs.
        Light blue line is the emission from trapped IXs and cavity modes, extracted by subtracting the baseline emission from the raw data.
        (b) Fitting of the trapped IXs (red peaks) and cavity modes (black peaks) by multiple peaks.
    }
\end{figure}

We observe superimposed emission peaks from delocalized IXs, moiré trapped IXs and high-Q cavity modes in the raw spectra.
For example, in Fig.~\ref{sf4} we present one raw spectrum (dark blue) recorded without magnetic field (using the setup in Fig.~\ref{sf3}(b)), excited by the 532-nm laser having a power of 0.6 $\mathrm{\mu W}$.
The broad baseline emission is from free and delocalized IXs.
The narrow and sharp peaks are from trapped IXs and cavity modes.
We subtract the baseline (gray) from the raw data to get the narrow and sharp emission from trapped IXs and cavity modes (light blue).

Then we extract properties of trapped IXs and cavity modes by a multi-peak fitting.
For the spectrometer used in the spectroscopy without magnetic field, the broadening is a convolution of Gaussian function with the width of 50 pm \cite{Nomura2010,2210.00150}.
Thereby, we use Voigt peak (convolution of Lorentz and Gaussian) to fit the PL spectra, as shown by the example in Fig.~\ref{sf4}(b).
For accuracy, we fit in the wavelength dimension and the Gaussian width of the Voigt peak is fixed at 50 pm \cite{Nomura2010,2210.00150}.
The real linewidth is extracted by the Lorentzian width of the Voigt peak and transformed to the energy dimension.
For example, the Voigt fitting of the emission from cavity mode M1 gives a Lorentzian width $108$ $\mathrm{\mu eV}$, corresponding to a Q-factor of $1.2\times 10^4$.
This Q-factor is further strengthened by the coherence time measurement later in Sec.~\ref{secs1d}.

For the magneto PL spectroscopy, the FWHM of mode M1 measured by the 0.3m spectrometer is $263 \pm 2$ $\mathrm{\mu eV}$.
Compared to the real value $108\ \mathrm{\mu eV}$, we obtain a broadening $\sim160$ $\mathrm{\mu eV}$ from the spectrometer.
This value is consistent to the broadening extracted from the linewidth of laser signal, which is $\sim 80$ pm as measured and $<$1 pm in reality.
Thereby, we simply use Lorentz fitting in the magneto PL spectroscopy for brevity, and decrease the raw linewidth by 160 $\mathrm{\mu eV}$ for the deconvolution.
We emphasize that, the lasing feature, e.g., Fig.~2(b) is the narrowing of cavity linewidth.
The existence or suppression of the narrowing linewidth is clearly observed from the relative variation of the extracted value, not affected by a constant convolution (broadening) in absolute.

\subsection{\label{secs1d} Cavity Q-Factor}

\begin{figure}
    \includegraphics[width=\linewidth]{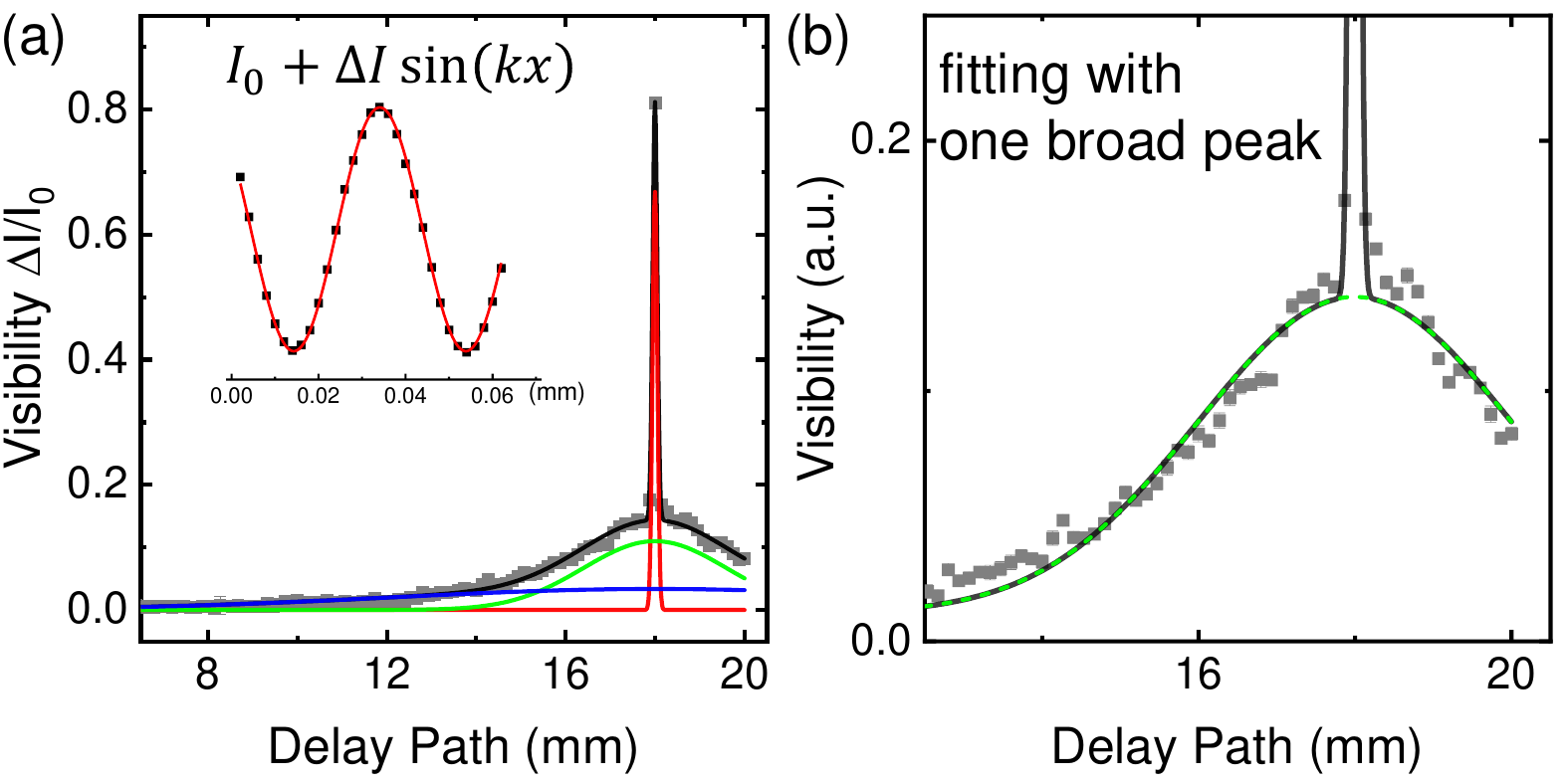}
    \caption{\label{sf5}
        Coherence time measurement.
        (a) The visibility of the interference (inset) with varying delay path, fitting by one sharp (red) and two broad (green and blue) Gaussian peaks.
        (b) Fitting with only one broad Gaussian peak (dashed green) cannot well describe the data.
    }
\end{figure}

In Fig.~\ref{sf5} we present the coherence time (length) of cavity mode M1 that further confirms the high Q-value.
This measurement is based on the interference with the phase differences induced by a delay line \cite{doi:10.1126/sciadv.aav4506}.
Due to the interference between the photons with and without delay, the collected photon count will vary periodically with the delay path.
We firstly vary the delay path $x$ in a small range such as shown by the inset in Fig.~\ref{sf5}(a).
The photon count $I$ with interference is then fitted by $I_0+\Delta I\sin\left(kx\right)$ where $I_0$ is the average intensity, $\Delta I$ is the amplitude of intensity variation, and $k$ reflects the photon wavelength.
We measure the visibility $\Delta I/I_0$ at a wide range of delay path.
The results are presented in Fig.~\ref{sf5}(a).
The Fourier transform of the cavity emission peak results in a Gaussian shape in the visibility with a width of $c\tau$, where $\tau$ is the coherence time and $c$ is the vacuum light speed.
We use multi Gaussian fitting to extract the coherence time \cite{doi:10.1126/sciadv.aav4506}.
The best fitting is achieved by three Gaussian peaks with the coherence time $\tau_1=39 \pm 3$ $\mathrm{ps}$, $\tau_2=10.7\pm 0.4$ $\mathrm{ps}$ and $\tau_3=355 \pm 9$ $\mathrm{fs}$ denoted by the color peaks in Fig.~\ref{sf5}(a).
Then the Q-factor is extracted as
\begin{eqnarray}
    \label{spe}
    Q=\frac{\lambda}{\Delta\lambda}=\frac{c\tau}{\lambda} \nonumber
\end{eqnarray}
where $\Delta\lambda$ is the linewidth and $\lambda$ is the wavelength.
The first broad Gaussian peak (blue) in Fig.~\ref{sf5}(a), extracted from the coherence time data, gives $\tau_1=39 \pm 3$ $\mathrm{ps}$ corresponding to $Q_1=1.25 \pm 0.08 \left(\times 10^4\right)$.
By comparison, the PL spectrum discussed in Sec. \ref{secs1c} gives $Q=1.2\times 10^4$ corresponding to $\tau=38$ $\mathrm{ps}$.
This perfect agreement between two measurements reveals that in our work, the cavity linewidth extracted from PL spectroscopy is accurate and correctly reflects the coherence time.
Therefore, we believe that the linewidth narrowing observed from PL spectra (Fig. 2b) demonstrates the increase of coherence time.

The second broad Gaussian peak (green) corresponds to $Q_2=3.4 \pm 0.1 \left(\times 10^3\right)$.
As shown in Fig.~\ref{sf4}(b), the detection wavelength range $900-950$ nm in this measurement involves multiple cavity modes.
Because the intensity of moiré IX is much weaker than the cavity emission, we suggest the broad peak with Q2 arising from another cavity mode.
The third sharp Gaussian peak (red) corresponds to $Q_3=114 \pm 3$.
This short lifetime emission is consistent to the broad base emission of free and delocalized IXs.

We note that in Fig.~\ref{sf5}(a) we use two broad Gaussian peaks (green and blue) in the fitting.
For comparison, in Fig.~\ref{sf5}(b) we plot the fitting with one narrow peak of delocalized IXs (not plotted) and only one broad Gaussian peak (green dashed line) of cavity mode.
As shown, this fitting cannot well describe the experimental data, supporting the multiple broad Gaussian peaks in Fig.~\ref{sf5}(a).
Indeed, since more cavity modes exist within the detection range, we suggest more broad Gaussian peaks exist in the visibility, but they cannot be extracted from the present data.
This measurement is limited by the time i.e., due to the high Q-factor, we need to trace a large range of delay path to obtain the broad Gaussian peaks from cavity modes.
The data in Fig.~\ref{sf5} is recorded at a high excitation power $10\ \mathrm{\mu W}$ but takes more than three hours.
Thereby, more data points in the measurement, or more measurements at lower excitation powers is infeasible at present.

\section{\label{secs2} Identifying Moiré Signatures}

\begin{figure}
    \includegraphics[width=\linewidth]{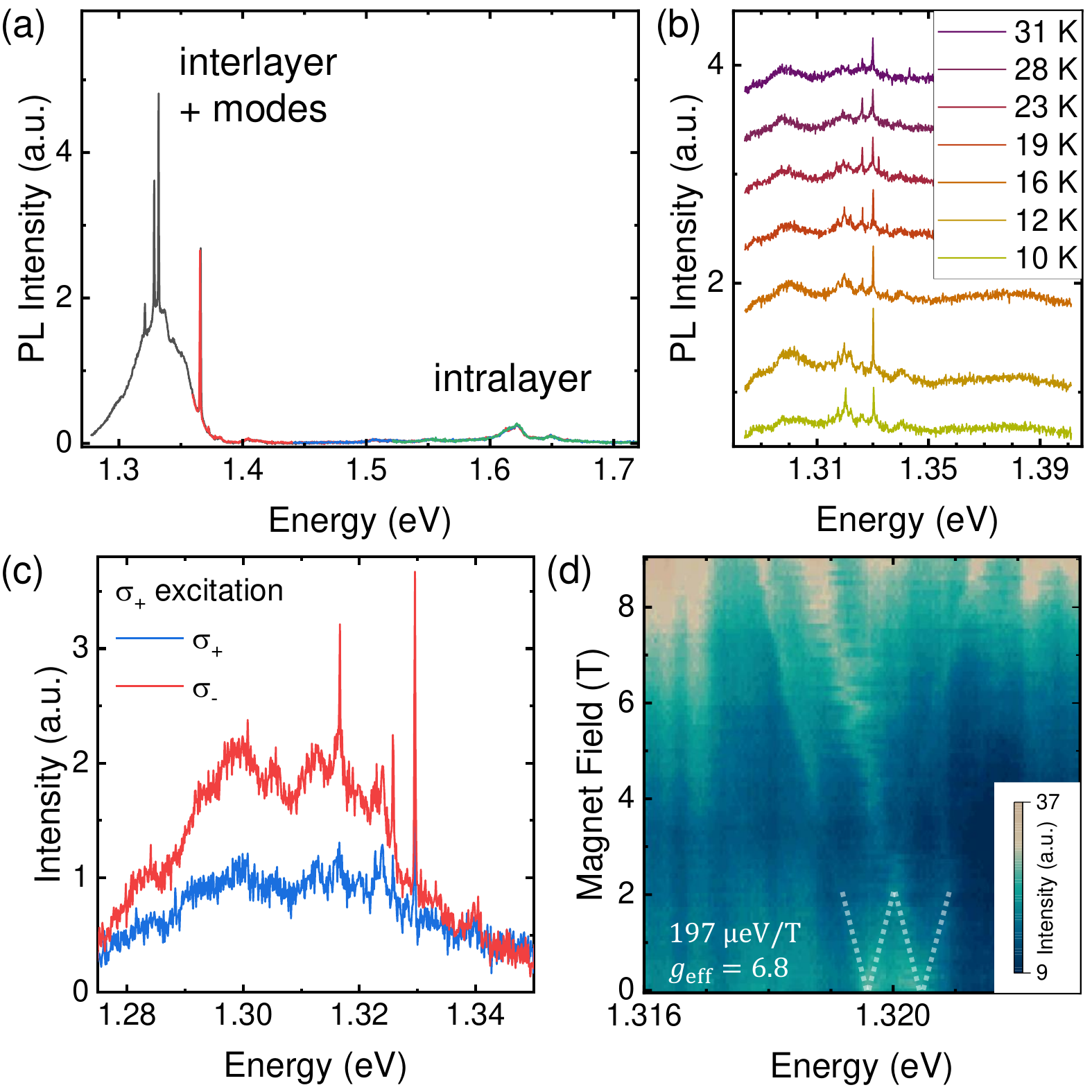}
    \caption{\label{sf6}
        Moiré signatures.
        (a) Full range spectrum recorded at the high power of $53\ \mathrm{\mu W}$.
        Four peaks within $1.60-1.67$ eV are the intralayer excitons.
        Peaks below 1.4 eV are the IXs and cavity modes.
        (b) Temperature-dependent PL spectra of the cavity.
        As temperature increases, the emission of moiré trapped IXs rapidly suppresses.
        (c) Negative circular polarization dependence are observed for both the trapped and delocalized IXs.
        Spectra in (a)-(c) are measured in the setup without magnets.
        (d) Zeeman shift and the similar $g_\mathrm{eff}$ around 6.8 are observed for all trapped IXs.
    }
\end{figure}

In Fig.~\ref{sf6}(a) we present the PL spectrum of the cavity recorded at an excitation power of $53\ \mathrm{\mu W}$.
The spectrum has four periods and each is recorded by the micro-PL setup (Fig.~\ref{sf3}(b)) but with the grating of 300 grooves per mm.
Peaks below 1.4 eV are from the IXs and cavity modes, as already discussed in the context of Fig.~\ref{sf4}.
In addition, the four peaks around $1.60-1.67$ eV are from the intralayer excitons, including the neutral exciton and trion from the monolayer MoSe$_2$ and WSe$_2$.

In Fig.~\ref{sf6}(b) we present the PL spectra recorded at $73\ \mathrm{nW}$ with the temperature $T$ varying from 10 to 31 K.
As shown at low $T$, we observe the narrow emission peaks from both the trapped IXs and the cavity modes.
As $T$ increases, the emission of moiré trapped IXs suppresses due to the transition to delocalized IXs \cite{Seyler2019,doi:10.1021/acsnano.0c08981,Mahdikhanysarvejahany2021}.
This suppression at high $T$ is one moiré signature \cite{Seyler2019}.
Meanwhile, the peaks from cavity modes are not suppressed, since the background emission feed the photons in cavity as discussed in the context of Fig.~3(a).

\begin{figure}
    \includegraphics[width=\linewidth]{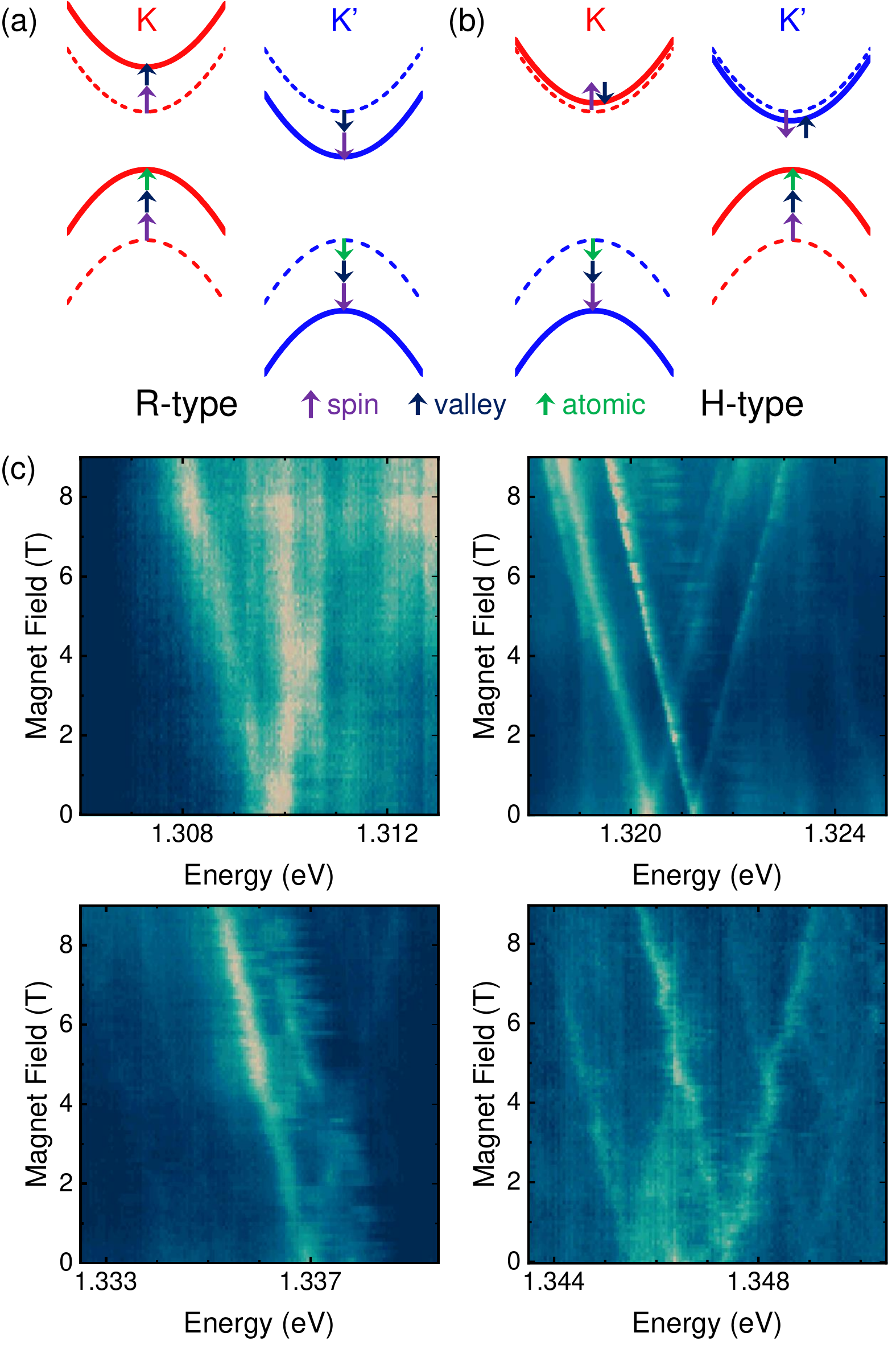}
    \caption{\label{sf7}
        Zeeman shift of IXs.
        (a) Schematic of Zeeman shift for R-type and (b) H-type heterobilayer TMDs.
        Dashed lines denote the valley without magnets, and solid lines denote the valley with magnetic field.
        The Zeeman shift of IXs includes contributions from the electron spin (purple arrow), the valley (dark blue arrow) and the atomic orbital (green arrow).
        (c) Magneto PL spectra measured recorded at 1.6 K.
        All trapped IXs exhibit the similar $g_\mathrm{eff}$ around 6.8.
        Such a small value corresponds to the R-type with a twist angle 2$^\circ$.
    }
\end{figure}

In Fig.~\ref{sf6}(c) we present the PL spectra excited by a $\sigma_+$ polarized laser and collected at $\sigma_+$/$\sigma_-$ channels.
For both the delocalized (broad baseline) and trapped (narrow peaks) IXs, the emission at $\sigma_-$ channel is much stronger than that at $\sigma_+$ channel.
This is the typical negative circular polarization in the R-type heterobilayer TMDs \cite{Seyler2019,Jiang2021,Hsu2018}.
Moreover, the similar Zeeman shift and effective g-factor $g_\mathrm{eff}\approx6.8$ are observed for all trapped IXs, such as the magneto PL spectra presented in Fig.~\ref{sf6}(d).
The small $g_\mathrm{eff}$ is due to that, when we apply the magnetic field to the R-type MoSe$_2$/WSe$_2$ heterobilayer, the Zeeman shift arising from electron spin and valley moment have the same direction in the conduction and valance band, as shown in Fig.~\ref{sf7}(a).
As a result, the corresponding shift in the conduction band is cancelled by the valance band, and the Zeeman shift of IXs only has contributions from the atomic orbitals, much smaller than the H-type heterobilayer shown in Fig.~\ref{sf7}(b).
We present several magneto PL spectra in Fig.~\ref{sf7}(c).
The temperature here is 1.6 K thereby the trapped IXs have narrower linewidth \cite{Mahdikhanysarvejahany2021}.
As shown, all the trapped IXs have a similar $g_\mathrm{eff}$ around 6.8.
The $g_\mathrm{eff}$ value and negative polarization reveal that the trapped IXs inherit the valley contrasting properties of R-type heterobilayer with a twist angle 2$^\circ$, featuring moiré potential in contrast to other randomly formed extrinsic traps \cite{Seyler2019,doi:10.1021/acs.nanolett.1c01215,Mahdikhanysarvejahany2022}.

\section{\label{secs3} Control Experiments}

\subsection{\label{secs3a} Green Laser Excitation}

\begin{figure}
    \includegraphics[width=\linewidth]{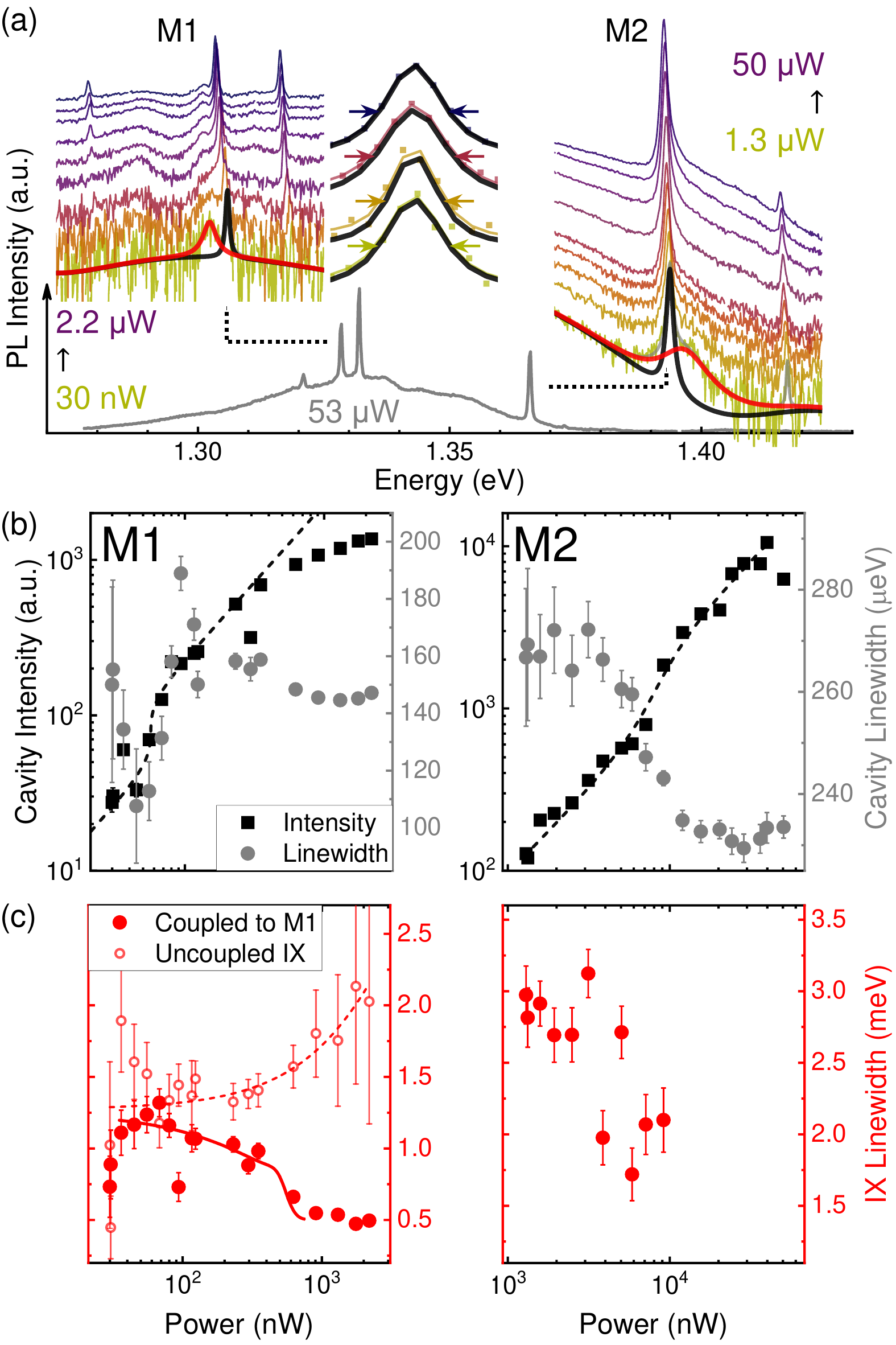}
    \caption{\label{sf8}
        Lasing under the green laser excitation.
        (a) Bottom gray spectrum is excited by a high power of $53\ \mathrm{\mu W}$, including the sharp peaks from cavity modes and the broad base emission of delocalized IXs.
        Insets are spectra (log scale) recorded at lower powers around the mode M1 and M2 (black peak).
        We also observed narrow emission peaks arising from moiré trapped IXs (red peaks).
        The variation of M1 linewidth is denoted by the four pairs of arrows.
        (b) The power-dependent intensity (black) and linewidth (gray) of two cavity modes, which have been presented in Fig. 4 in the main paper.
        (d) The linewidth of IXs coupled to the cavity mode also narrows with the increasing excitation power, as a distinct feature for the lasing \cite{Nomura2010}.
        In contrast, the linewidth of uncoupled IX (e.g., hollow dataset) increases.
        Note: the trapped IX near M2 is difficult to fit at high excitation power $>10$ $\mathrm{\mu W}$.
    }
\end{figure}

We also implement PL spectroscopy without magnets excited by the 532-nm cw-laser (Fig.~\ref{sf3}(b)).
Typical PL spectra recorded at 10 K are presented in Fig.~\ref{sf8}(a).
At a high excitation power $53\ \mathrm{\mu W}$ (bottom gray spectrum), we observe the broad emission from an ensemble of free and delocalized IXs ($1.28-1.37$ eV) \cite{Rivera2015}, along with several sharp peaks from cavity modes.
As the excitation power decreases, the moiré potential exceeds the excitation level, and thereby, the delocalized IXs transit to moiré trapped IXs \cite{doi:10.1021/acsnano.0c08981,Mahdikhanysarvejahany2021}.
We observed narrow emission peaks from these moiré trapped IXs at low powers, and some of them (red peak) are near resonant to the cavity (black peak) as presented in the insets of Fig.~\ref{sf8}(b).
We fit these peaks using the method discussed in Sec.~\ref{secs1c}.

The power-dependent intensity and linewidth of two cavity modes M1-2 are presented in Fig.~\ref{sf8}(b) (also Fig. 4 in the main paper).
The Q-factor of mode M1 is 12500 as discussed in Sec.~\ref{secs1d}, and the Q-factor of mode M2 is around 6000 calculated from the Lorentzian width.
The lasing features under the green laser excitation have been discussed in the context of Fig.~4.
For example, the re-broadening of cavity linewidth is explained by the coupling between intensity and phase noises \cite{Wu2015,doi:10.1002/lpor.201000039,doi:10.1063/1.106693,doi:10.1063/1.105443}, and the delocalized IXs contribute to the phase noises since they have lower purity and more fluctuations \cite{PhysRevB.86.161412,Mahdikhanysarvejahany2021}.
As shown in Fig.~\ref{sf8}(a), the mode M1 is at 1.33 eV, the peak of broad baseline, thereby have more photons contributed from delocalized IXs.
In contrast, the mode M2 at 1.36 eV have fewer photons from delocalized IXs since they are spectrally detuned.
As a result, the re-broadening is clearly observed for M1 but not for M2.

Since both the broad base emission (free and delocalized IXs) and the narrow emission (moiré trapped IXs) contribute to the photons confined in the cavity, and we cannot control the detuning in this setup, we probe the lasing of trapped IXs by their linewidth.
Without the coupling to cavity modes, the linewidth of trapped IXs is broadened by the increasing excitation power due to the increasing delocalization effects \cite{Seyler2019,Mahdikhanysarvejahany2021}.
This broadening is observed for the trapped IXs uncoupled (detuned) to the cavity mode, e.g., the hollow dataset in Fig.~\ref{sf8}(c).
In contrast, the lasing oscillation in the system will narrow not only the cavity mode but also the exciton \cite{Nomura2010,PhysRevB.81.033309}.
We observe this narrowing in the trapped IXs coupled (resonant) to the cavity mode presented by the solid dataset in Fig.~\ref{sf8}(c), supporting the lasing of moiré trapped IXs excited by the green laser.

\subsection{\label{secs3b} Coupling to Multiple Moiré IX Lines}

\begin{figure}
    \includegraphics[width=\linewidth]{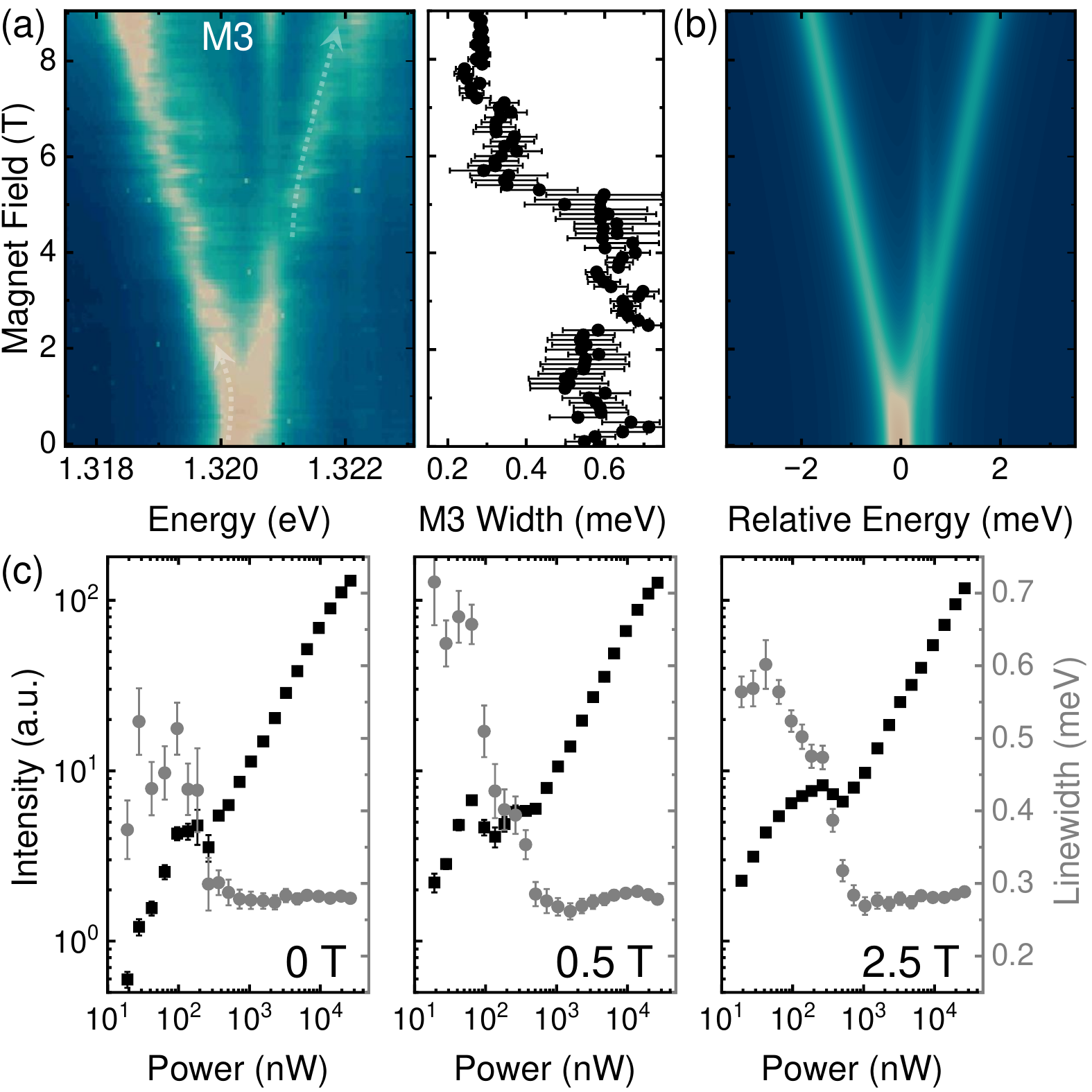}
    \caption{\label{sf9}
        Coupling to multiple moiré IX lines.
        (a) Magneto (detuning) dependent PL spectra at low excitation power 80 nW.
        The IX peak involves multiple IX lines (details in Fig.~\ref{sf7}) thereby has larger coupling strength $g'$.
        The nonlinear energy shifts indicate the strong coupling regime.
        Right panel shows the linewidth of cavity mode M3.
        (b) Calculation by a brief model.
        (c) Lasing features at three different magnetic field 0, 0.5, and 2.5 T.
        Due to the large $g'$, the linewidth narrowing is very pronounced.
    }
\end{figure}

In Fig.~\ref{sf9} we present another cavity mode M3 which couples to multiple moiré IX lines.
The magneto PL spectra at 10 K are presented in Fig.~\ref{sf9}(a).
Compared to the spectra of moiré IXs at 1.6 K in Fig.~\ref{sf7}, we obtain that the broad IX peak at 10 K in Fig.~\ref{sf9}(a) involves at least two transition lines, and thereby, the effective coupling strength is at least $g'=\sqrt{2}g$.
As denoted by the arrows, the emission peaks exhibit nonlinear energy shifts.
Meanwhile, the cavity linewidth as shown in the right panel of Fig.~\ref{sf9}(a), broadens from 280 $\mathrm{\mu eV}$ at 9 T (strongly detuned) to around 600 $\mathrm{\mu eV}$ at $<$4 T (near resonant to IX lines).
These features indicate the strong coupling regime, which is consistent to the larger effect coupling strength for multiple moiré IX lines.
In contrast, weak coupling regime (effective coupling strength $<$ 100 $\mathrm{\mu eV}$) cannot lead to such significant nonlinear shift and linewidth broadening.

It is non-trivial to repeat Fig.~\ref{sf9}(a) in calculation, because the IX peak might be an ensemble of multiple lines at slightly different energies.
Moreover, complex phonon effects such as the scattering between different states and the condensation to low energy levels exist in the system \cite{PhysRevResearch.2.042044}.
These specific details are not the scope of this work.
We use a brief model with one collective emitter and the effect coupling strength $g'=\sqrt{3}g$, without phonon effects, to breifly describe the strong coupling as presented in Fig.~\ref{sf9}(b).
Nonetheless, we clearly observe the much more pronounced lasing features including the threshold behavior and the linewidth narrowing in Fig.~\ref{sf9}(c), compared to the case of single moiré IX line in Fig. 2.
These stronger lasing features arise from the large $g'$.

\section{\label{secs4} Theoretical Calculation}

\subsection{\label{secs4a} Coupling Strength}

\begin{figure}
    \includegraphics[width=\linewidth]{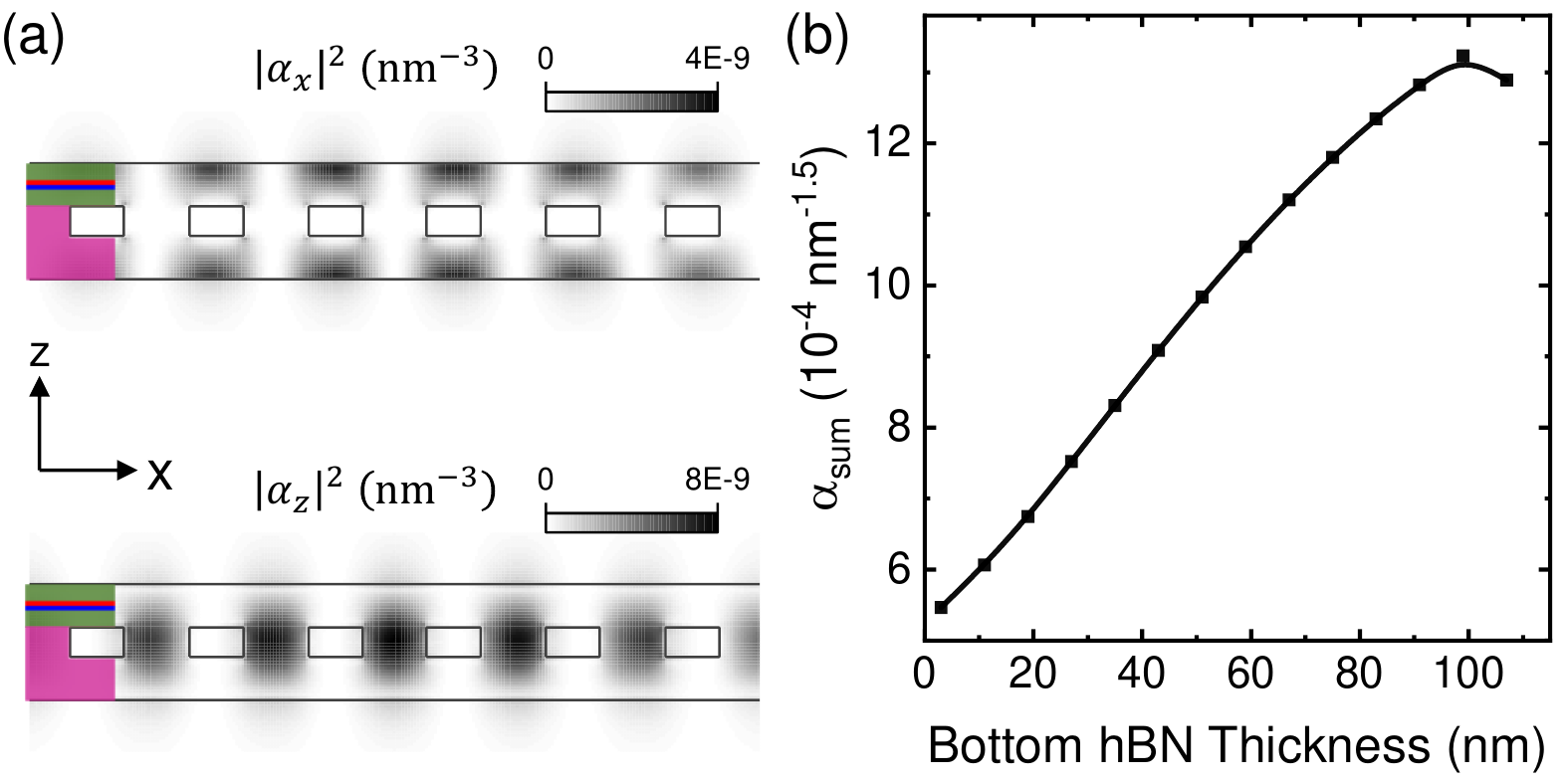}
    \caption{\label{sf10}
        Coupling strength in theory.
        (a) The cavity electric field.
        The in-plane $\alpha_x$ and out-plane $\alpha_z$ have the same magnitude.
        (b) The integral $\alpha_{\mathrm{sum}}$ vs. the thickness of bottom hBN.
        We use 40-nm thick bottom hBN to trade off to the feasibility in stacking.
    }
\end{figure}

We calculate the theoretical coupling strength between the cavity mode and one IX line by integrating all moiré superlattice sites in the cavity electric field as shown in Fig.~3(a).
We use the Tavis–Cummings model based on the fundamental physical features of moiré IXs.
Firstly, moiré IXs are dominated by the zero-phonon emission, whilst the phonon sideband at low temperature is weak.
In this case, the coupling to the cavity mode follows the Tavis–Cummings model, for which the exciton directly couples to the photons.
In contrast, for some localized emitters having weak zero-phonon emission whilst strong phonon sideband \cite{2210.00150}, their coupling to the cavity mode is mainly assisted by the phonon-induced processes thus the Tavis–Cummings model might be invalid.
Secondly, moiré IXs are localized in the moiré sites, and thereby, the local cavity electric field is approximately uniform and the interaction to the photon locally.
Therefore, we calculate the coupling strength based on the dipole moment.
In contrast, when applied with free excitons in 2D materials, the excitons will move through different positions in spatial having different electric field, and thereby, the coupling enters the nonlocal regime \cite{PhysRevLett.128.237403}.
As shown, the fundamental physics of moiré IXs determines the exciton-photon coupling and the lasing reported in this work.

In our calculation, the trapped IXs couple to the in-plane cavity electric field, since the optical dipole moment of IX is previously found to be in-plane \cite{PhysRevB.105.035417}.
As shown in Fig.~\ref{sf10}(a), the in-plane field $\alpha_x$ is indeed maximum around the top of the cavity structure, that means to maximize the coupling we need a thin (thick) top (bottom) hBN.
However, if we firstly pick up a thin top hBN using the stamp (Fig.~\ref{sf1}(c)), we will very likely induce wrinkles to this thin hBN, since the stamp will be deformed by the varying temperature during the stacking.
Therefore, to trade off between the coupling and the feasibility of stacking, we use 70 (40) nm thickness for the top (bottom) hBN.

In addition, the moiré trapped IX might have optical dipole approximation in a direction different to the free IX, and thereby, might be not in-plane.
This is the second reason for using 70 nm thick top hBN.
As shown in Fig.~\ref{sf10}(a), the in- and out-plane electric field have the same magnitude around the middle of hBN encapsulation.
This means that even if the moiré trapped IX has different dipole directions, their coupling strength to the cavity mode will be similar to the theory.

We note that the specific details of the calculation is limited by many factors.
The refractive index of hBN is not well known yet, thus for brevity we use the value of 2.0 in the calculation \cite{PhysRevLett.128.237403}.
This might introduce a minor difference between the calculated and real cavity electric field.
The thickness of hBN flakes are judged by their color contrast in the optical image.
This is a convenient method to quickly extract the approximate thickness for a huge amount of flakes, but has an inaccuracy $\sim20$ nm.
This difference in hBN thickness will shift the cavity mode \cite{10.1515/nanoph-2023-0347}, and moreover, affect the coupling strength $g$ as presented in Fig.~\ref{sf10}(b).
Meanwhile, the coupling strength might also be affected if some moiré sites are disordered by defects or contamination.
Nonetheless, we emphasize that the minor factors might affect the specific details in the calculation but not the magnitude of theoretical results.

\subsection{\label{secs4b} Lasing in Master Equation Theory}

\begin{figure}
    \includegraphics[width=\linewidth]{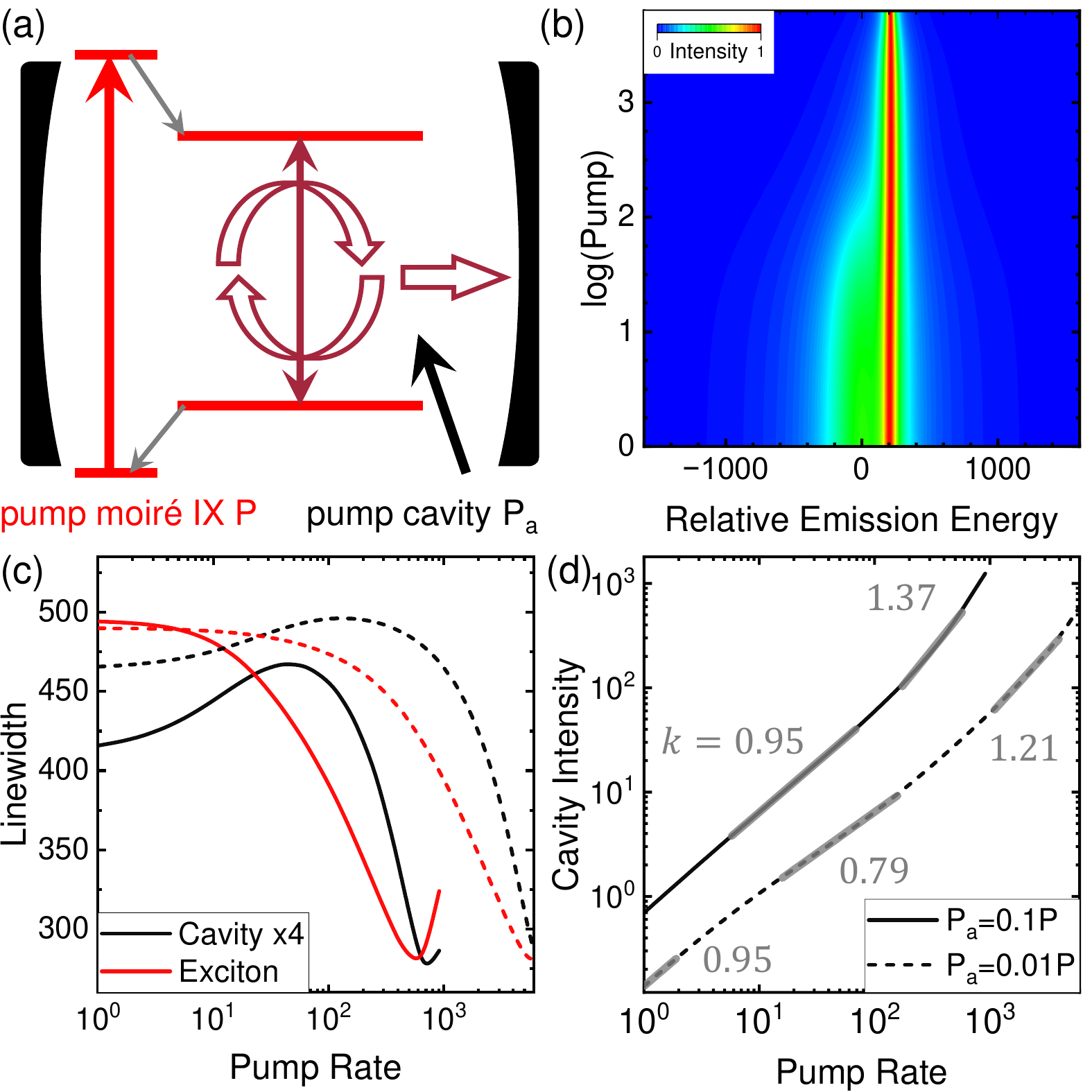}
    \caption{\label{sf11}
        Master equation theory.
        (a) Schematic of the system, including a cavity mode and a 4-level emitter.
        The system involves the coherent cavity-moiré-IX coupling and the incoherent feeding of cavity photons arising from the background emission.
        (b) Normalized emission spectra calculated with the pump $P=1-10^4$.
        (c) The linewidth of cavity and exciton peaks.
        (d) The emission intensity of cavity.
        Solid lines in (c)(d) are calculated with $P_a=0.1P$ and dashed lines are calculated with $P_a=0.01P$.
    }
\end{figure}

We use the master equation theory to calculate emission of the coupling system with a four-level emitter and a single cavity mode \cite{Nomura2010}, as depicted in Fig.~\ref{sf11}(a).
The system Hamiltonian is
\begin{eqnarray}
    % \label{eqH}
    \textit{H} &=& \hbar\omega_{a}a^{+}a + \sum_{i=1}^{4}\hbar\omega_{i}\sigma_{i,i}  + \hbar g\left(\sigma_{2,3}a^{+} + \sigma_{3,2}a\right) \nonumber
\end{eqnarray}
where $\sigma_{i,j}=|i \rangle \langle j|$ is the Dirac operator for the four level emitter with levels $1-4$, and $a^{+}$/$a$ are ladder operators for photons in the cavity mode.
Level 1 of the emitter is the ground state with the energy $\omega_1=0$.
Level 2 has the energy $\omega_2=\omega_{12}$ where $\omega_{12}$ is the $2 \rightarrow 1$ transition energy.
Similarly, $\omega_3=\omega_{12}+\omega_{23}$ and $\omega_4=\omega_{12}+\omega_{23}+\omega_{34}$ where $\omega_{23}$ ($\omega_{34}$) is the $3 \rightarrow 2$ ($4 \rightarrow 3$) transition energy.
$\omega_a$ is the energy of photon, and $g$ is the coupling strength between the cavity mode and the radiative $3 \rightarrow 2$ transition.

The master equation is given by
\begin{eqnarray}
    % \label{master}
    \frac{d}{dt}\rho&=&-\frac{i}{\hbar}[H,\rho]+\sum_{n}{\cal L}(c_n) \nonumber
\end{eqnarray}
where $\rho$ is the density matrix of the system with maximum $N_a$ photons allowed, $H$ is the Hamiltonian defined above, and the Liouvillian superoperator
\begin{eqnarray}
    % \label{decay}
    {\cal L}(c_n)=1/2\left(2c_n\rho c_n^+-\rho c_n^+c_n-c_n^+c_n\rho \right) \ \ \ \ \ \ \ \ \nonumber \\
    c_{n}=\lbrace \sqrt{\gamma_a}a,\sqrt{\gamma_{i-1,i}}\sigma_{i-1,i},\sqrt{P_a}a^{+},\sqrt{P}\sigma_{4,1}\rbrace,\ i\in[2,4] \nonumber
\end{eqnarray}
describes Markovian processes corresponding to the decay of photons with the rate $\gamma_a$, the decay between the four levels of emitter with the rate $\gamma_{i-1,i}$, the pump of photons with the rate $P_a$ and the pump of emitter with rate $P$, respectively.
For brevity, we use energy unit for all the parameters (e.g. $\omega_{a}$ is actually $\hbar\omega_{a}$) and omit the unit $\mathrm{\mu eV}$.
The emission spectrum is calculated from the steady state of the master equation.

In Fig.~\ref{sf11}, we present typical results calculated with $\omega_a=200$, $\omega_{12}=-30000$ and $\omega_{43}=-70000$ relative to the emission energy of exciton set to $\omega_{23}=0$.
Based on the experimental observation, we set $\gamma_a=100$, $g=78$, $\gamma_{2,3}=495$ and $\gamma_{1,2}=5$.
The pump related parameters of emitter are set as $\gamma_{3,4}=0.5P$ with the pump $P$ varying between $1-10^4$.
In addition, background emission, e.g., the delocalized IXs and exciton-exciton interaction, feeds photons in the cavity \cite{Nomura2010}.
We set this pump of cavity as $P_a=0.01P$.
Due to the limitation of the memory of our computer, we set at maximum seven photons in the cavity mode ($N_a=7$).
We note that the coupling between intensity and phase noise \cite{Wu2015,doi:10.1002/lpor.201000039,doi:10.1063/1.106693,doi:10.1063/1.105443} is also not included in this system.
Thereby, the calculation results with large $P$ are slightly affected.
Despite minor differences owing to these limitations, the theoretical calculation generally reproduces our observation, as discussed in the main paper.

\begin{figure}
    \includegraphics[width=\linewidth]{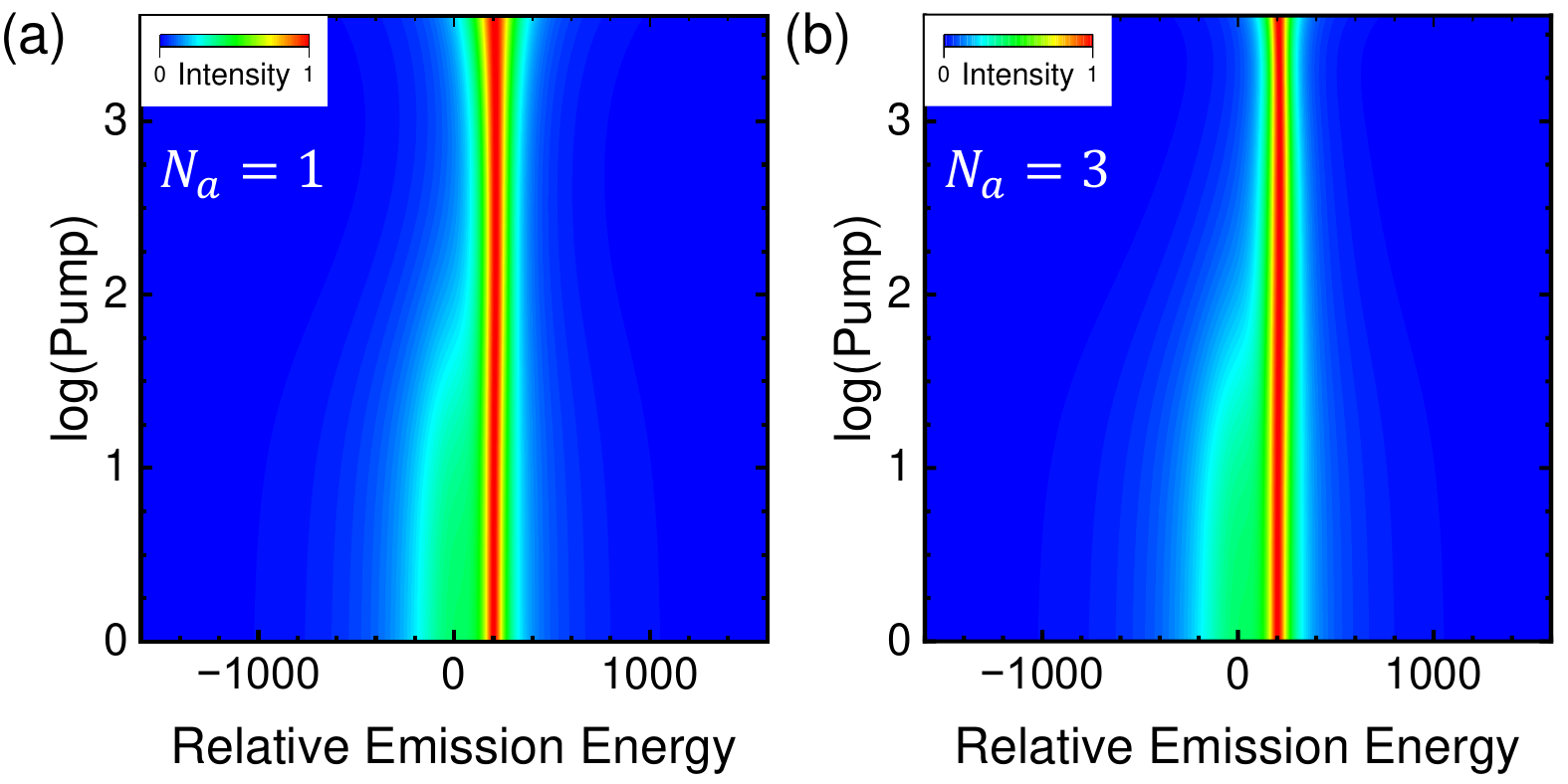}
    \caption{\label{sf12}
        Normalized spectra calculated with different number of photons $N_a$.
        (a) Lasing, i.e., the cavity linewidth narrowing, will never occur with only one photon.
        (b) For $N_a=3$, calculation with the pump rate $>10^3$ becomes inaccuracy, because such high pump rate generates more than 3 photons in the cavity.
        In contrast, $N_a=7$ in Fig.~\ref{sf11} is enough for the calculation.
    }
\end{figure}

Normalized emission spectra of the system are shown in Fig.~\ref{sf11}(b).
The peak at 0 is from the $3 \rightarrow 2$ transition of exciton, and the sharp peak at 200 is from the cavity.
As the pump $P$ increases, the ratio between the intensity of exciton and cavity is suppressed, corresponding to the increasing stimulated emission.
The linewidth of two peaks are presented in Fig.~\ref{sf11}(c), and the cavity intensity is presented in Fig.~\ref{sf11}(d).
The solid lines are calculated with $P_a=0.01P$, and the dashed lines are in addition calculated with $P_a=0.1P$ for comparison.
We note that the narrowing of cavity linewidth in (c) starts at the sublinearity, rather not the superlinearity of L-L curve in (d) \cite{Nomura2010,Ritter:10}.
Meanwhile, as presented in Fig.~\ref{sf11}(c)(d), the narrowing rate of cavity linewidth and the superlinearity $k$ of cavity intensity both increase with $P_a$.
This provides an explanation for the larger superlinearity under the green laser excitation as discussed in the main paper, because the green laser generate more background emissions which results in the larger $P_a$.

Generally, the limitation $N_a=7$ is enough within the pump rate range in our calculation.
We present the case of $N_a=1$ and $N_a=3$ in Fig.~\ref{sf12} for comparison. 
As shown in Fig.~\ref{sf12}(a), if only 1 photon is allowed in the cavity, the lasing, i.e., the cavity linewidth narrowing, will never occur.
This agrees well to the fact that, the nanocavity lasing essentially means the number of photons in the cavity larger than one \cite{PhysRevA.50.1675}.
For $N_a=3$, we observe the linewidth narrowing when the pump rate exceeds $10^2$.
However, as the pump rate further increases ($>10^3$), the cavity linewidth increases like the case of $N_a=1$ in Fig.~\ref{sf12}(a).
This is because such high pump rate generates more than 3 photons in the cavity, thus the limitation of $N_a=3$ leads to the inaccurate results.
In contrast, we do not observe such pump-induced dephasing (linewidth broadening) for $N_a=7$ in Fig.~\ref{sf11}, which means this value is enough for the calculation in this work.

\end{document}